%% file: main.tex
\documentclass[aps,prd,reprint,nofootinbib,amsmath,amssymb,longbibliography,superscriptaddress]{revtex4-2}

\usepackage[T1]{fontenc}
\usepackage[utf8]{inputenc}
\usepackage{graphicx}
\usepackage{xcolor}
\usepackage{hyperref}
\hypersetup{
    colorlinks=true,
    linkcolor=red!60!black,
    citecolor=green!50!black,
    urlcolor=blue!70!black
}

\usepackage{dsfont}
\usepackage{tikz}
\usepackage{tensor}
\usepackage{dirtytalk}
\usepackage{empheq}

\usepackage{calligra}
\DeclareMathAlphabet{\mathcalligra}{T1}{calligra}{m}{n}
\DeclareFontShape{T1}{calligra}{m}{n}{<->s*[2.6]callig15}{}

\newcommand{\scriptr}{\mathcalligra{r}\,}

\begin{document}

\title{A note on Holographic Renormalization for Polarized IKKT}
\author{Matteo Guglielmetti}
\affiliation{SISSA, Trieste, Italy}
\date{\today}

\begin{abstract}
We construct a boundary counterterm for the on-shell action of type IIB supergravity
solutions dual to vacua of the polarized IKKT matrix model. The construction is based
on the variational principle, with off-shell fluctuations whose fall-off conditions
are based on the asymptotic equations of motion. The renormalized on-shell action
of the supersymmetric saddles reproduces the matrix-model result once we pass to the
ensemble of fixed D1-brane charges, which we implement off-shell by dualizing the RR
two-form potential.
\end{abstract}

\maketitle

\section{Introduction}
\label{sec:introduction}
\input{Introduction}

\section{Boundary Data and Fall-Off Conditions}
\label{sec:data}
\input{Data}

\section{Counterterm from Stationarity}
\label{sec:ct}
\input{Counterterm}

\section{The Electrostatic Family of BPS solutions}
\label{sec:electrostatics}
\input{electrostatics}

\section{Canonical Ensemble from Dualization}
\label{sec:canonical}
\input{canonical}

\pagebreak

\section{Conclusion}
\label{sec:conclusion}
\input{conclusion}

\begin{acknowledgments}
I am especially grateful to Jo\~ao Penedones and Antoine Vuignier for inviting me
to work on this problem and for useful discussions. I am grateful to Gabriel Cuomo and Giulio Bonelli for useful discussions and reading of the manuscript. I would also like to thank
Francesco Benini and Monica Guica for comments.
I acknowledge the use of ChatGPT (OpenAI) and Claude (Anthropic) to assist with
technical calculations and proofreading, and in the exploratory phases of this work.
\end{acknowledgments}

\appendix

\input{electro_app}

\bibliography{bib}

\end{document}

%% file: Introduction.tex
The polarized IKKT model is a supersymmetric matrix model that provides an instance of gauge / gravity duality \cite{hartnoll2024polarisedikktmatrixmodel}. In fact, it was shown in \cite{komatsu2025einsteingravitymatrixintegral,komatsu2025einsteingravitymatrixintegral2} that there exists a holographic correspondence between supersymmetric vacua and $1/2$-BPS saddles of type IIB supergravity with a background of $D$-instantons. In more detail, the bosonic sector of the IKKT model \cite{Ishibashi:1996xs} consists of a ten-dimensional vector of $\mathfrak{u}(N)$-valued fields $X^M$ and polarization (a supersymmetric mass deformation) breaks its $SO(10)$ symmetry explicitly down to $SO(3)\times SO(7)$ \cite{Bonelli_2002}. In the supersymmetric vacua, the $SO(7)$ vector vanishes, $X^m=0$, while the $SO(3)$ triplet $X^i$ takes the form of a \emph{fuzzy sphere geometry}: its components are proportional to generators of $SU(2)$, $X^i\sim J^i$, in a decomposable representation with irreducible blocks having dimensionality $N_s$, each appearing with multiplicity $n_s$. In particular, $N=\sum_s n_s N_s$. The dual geometries inherit the $SO(3)\times SO(7)$ symmetry as solutions to Euclidean type IIB supergravity, truncated to the $F_5=0$ sector, with $16$ supercharges. Such solutions are presented and discussed in \cite{komatsu2025einsteingravitymatrixintegral}. They take the form of a warped product $\Sigma\times S^2\times S^6$, with loci where the spheres shrink. They are fully specified by a scalar potential $V$ that solves Laplace's equation on $\Sigma\times S^2$ with $SO(3)$ symmetry.

Comparing on-shell actions under the holographic correspondence, one faces the issue of having to renormalize the IR divergences of the on-shell action in an asymptotically flat spacetime. In asymptotically $AdS$ spacetimes the field equations, solved order by order in the cutoff parameter, allow one to write the divergent terms as local functions of the sources. This recursive structure allows the construction of local counterterms valid for arbitrary source profiles at the conformal boundary. This is not the case for asymptotically flat spacetimes, where the equations of motion yield differential, rather than algebraic, constraints in the expansions. As a consequence, the sources are not arbitrary functions on $S^9$, and in general the recursion used in $AdS$ to construct the counterterm involves here the inversion of differential operators. The standard $AdS$ procedure for constructing universal local counterterms therefore does not extend directly to the asymptotically flat setting \cite{Skenderis_2002}.

Nevertheless, with the sources fixed to those of the polarized IKKT gravity duals, we construct local counterterms, intrinsic to the cutoff boundary, that reabsorb all divergences of the on-shell action for any saddle with $SO(3)\times SO(7)$ symmetry. However, distinct counterterms can cancel the same divergences while differing by finite terms. To resolve this ambiguity, we require holographic renormalization not only to cancel divergences but also to pose a well-defined variational problem in addition to a finite action. Concretely, the renormalized action must be stationary on solutions under all admissible off-shell variations at fixed sources \cite{Kanitscheider_2008,Papadimitriou_2010,Mann_2006,Mann_2010,Detournay_2014,dei2025onshellgravityactionlinear}. For variations of compact support, stationarity follows directly from the bulk equations of motion. Instead, fluctuations extending to infinity produce asymptotic boundary terms, which must be cancelled by the counterterms. We use this requirement to fix the response-dependent finite contribution to the action, up to terms depending only on the sources. The outcome of this work is a local, intrinsic counterterm that satisfies the finiteness and stationarity requirements on any $SO(3)\times SO(7)$ saddle with sources fixed by the polarized IKKT model.

\paragraph{Outline} We start in \autoref{sec:data} by carrying out a systematic analysis of the constraints that equations of motion impose on the asymptotic expansions of fields within an $SO(3)\times SO(7)$ ansatz. In doing so, one must specify the sources, defined as the coefficients of the growing modes. We will find which terms are forced and which are left free, the \emph{responses}, which change from saddle to saddle. We will also find the rates of fall-off of the solution and we will state those we assume for the off-shell fluctuations. In \autoref{sec:ct} we compute the bare regulated boundary contribution to the on-shell action and we present three distinct candidate counterterms that one could use to cancel the diverging part. The finite part differs and is dependent on the responses, making the three candidates not equivalent. We then show that the stationarity of the action selects a unique linear combination of the three, removing the ambiguity. In \autoref{sec:electrostatics} we briefly introduce the family of BPS solutions dual to the polarized IKKT vacua and compute its renormalized on-shell action. Finally, in \autoref{sec:canonical}, we change the ensemble to the canonical one, fixing $(n_s,N_s)$ through the choice of topological sector in the supergravity path integral. We end by showing a match with the matrix model on-shell action.

\paragraph{\textbf{Note added}} While this work was being finalized, \cite{najafian2026compactholographypolarizedikkt} appeared, obtaining the same matching but with a counterterm that differs from ours in the matter sector and whose stationarity was established only for fluctuations within the supersymmetric (electrostatic) family.

\paragraph{Notation.}
Before starting, let us fix some of the notation that will be recurrent. We work in Euclidean signature and in Einstein frame, with $2\kappa^2=(2\pi)^7g_s^2\alpha'^4$. Actions are denoted by $\mathcal{S}$ and their on-shell values by $I$. We only consider bosonic terms, leaving fermions implicit (see \hyperref[foot:other]{footnote 11}).
\begin{itemize}
\item \emph{Coordinates.} $x^M$, $M=1,\dots,10$, are asymptotically Cartesian harmonic coordinates, 
\begin{align} \label{eq:harmonic_gauge}
    \square_gx^M=0,
\end{align}
split into $x^i$ ($i=1,2,3$) and $x^m$ ($m=4,\dots,10$). We set $r^2=x^Mx_M$, $z=|x^i|=ru$ and $\rho=|x^m|=r\sqrt{1-u^2}$, with $0\leq u\leq 1$. The cutoff surface is $\partial\mathcal{M}_R=\{r=R\}$, with induced metric $h$, scalar curvature $R_h$ and outward unit normal $n$. Remainders of expansions are given in little-$o$ or big-$O$ notation with respect to $1/r$.

\item \emph{Fields.} In the $F_5=0$ truncated sector, the dynamical fields are the axion $\chi$, RR potential $C_2$, B-field $B_2$, metric $g_{MN}$ and dilaton $\phi$. The bosonic sector action reads
\begin{align}
\begin{aligned}
    \mathcal{S}_{\text{IIB}}=&-\frac{1}{2\kappa^2}\int_{\mathcal{M}_R}*_g\Big[R-\frac{1}{2}\left|d\phi\right|^2-\frac{1}{2}e^{2\phi}\left|d\chi\right|^2\\
    &\hspace{70pt}-\frac{1}{2}e^{-\phi}\left|H_3\right|^2-\frac{1}{2}e^\phi\left|F_3\right|^2 \Big]
\end{aligned}
\end{align}
On the contour of the solution, the real fields are
\begin{align} \label{eq:field_basis}
\begin{aligned}
    &y=e^\phi\quad\tau=e^{-\phi}+\text{i}\chi\\ &C_2=\text{i}c_2\quad H_3=dB_2\\
    &F_3=dC_2-\chi H_3\quad G_3=dc_2+\tau H_3.
\end{aligned}
\end{align}
Under $SO(3)\times SO(7)$, $c_2=c\,\omega_2$ and $B_2=b\,\omega_2$, where $\omega_p$ is the volume form of the round unit $S^p$ and $\Omega_p$ its volume. $\widetilde F_7$ and $\widetilde H_7$ are the electric seven-forms measuring $D1$ and $F1$ charges, \begin{align} \label{eq:stress_tensors}
\begin{aligned}
    -\text{i}\widetilde F_7&=*\left(yG_3-H_3\right),\\
    \widetilde H_7&=*\left[\left(1-y\tau\right)G_3+\tau H_3\right]
\end{aligned}
\end{align}

\item \emph{Forms.} $\langle A_p,B_p\rangle=\frac{1}{p!}A_{M_1\cdots M_p}B^{M_1\cdots M_p}$ and $|A_p|^2=\langle A_p,A_p\rangle$, with no complex conjugation. $*_g$ and $*_h$ are the Hodge duals of $g$ and $h$, $d_h$ is the exterior derivative on $\partial\mathcal{M}_R$, and $\iota_n$ is the interior product with $n$.

\end{itemize}

%% file: Data.tex
In this first section, we study the constraints on the asymptotic expansions of the matter fields and the metric that come from the equations of motion. After having specified the boundary conditions determined by a set of growing modes, the sources, we will be interested in studying the order of the first subleading corrections that are either forced by the sources or left free in the asymptotic analysis (which we refer to as responses).

In the field basis \eqref{eq:field_basis}, the bosonic matter action of type IIB supergravity takes the following form,
\begin{align} \label{eq:matter_action}
\begin{aligned}
    \mathcal{S}^{m}_{\text{IIB}}=\frac{1}{2\kappa^2}\int_{\mathcal{M}_R}&*_g\Big[-
\langle dy, d\tau\rangle-\frac{y^2}{2}|d\tau|^2
\\&\qquad\quad+\langle H_3,G_3\rangle-\frac y2|G_3|^2\Big]
\end{aligned}
\end{align}
The ansatz for the solutions in consideration is that of a warped product $\Sigma\times S^2\times S^6$ given by fibering an $S^2$ and $S^6$ over a two-dimensional base with a degeneration locus where they can shrink. Using $(r,u)$ as coordinates on the base, the ansatz has the explicit form
\begin{align} \label{eq:ansatz}
\begin{aligned}
    g&=ds^2_\Sigma+r_2(r,u)^2d\Omega_2^2+r_6(r,u)^2d\Omega_6^2\\[5pt]
    c_2&=c(r,u)\,\omega_2,\quad
    B_2=b(r,u)\,\omega_2\\[5pt]
    y&=y(r,u),\quad \tau=\tau(r,u).
\end{aligned}
\end{align}
Regularity of the geometry constrains the behavior of the $r_i^2$ functions, of the base metric and of the potentials near their zeros, as studied in \cite{komatsu2025einsteingravitymatrixintegral}. We will return to this in \autoref{sec:electrostatics}; here, we are instead interested in the asymptotic behavior and we assume regularity.

\begin{figure}[t]
    \centering
    \begin{tikzpicture}[scale=1.3]
  \draw[thick] (0,0) -- (2,0);
  \fill (0,0) circle (1.2pt) (2,0) circle (1.2pt);
  \node[below] at (0,-0.1)   {$u=0$};
  \node[below] at (1,-0.1) {$u$};
  \node[below] at (2,-0.1)   {$u=1$};
  \node[above] at (0,0.1)   {$S^6$};
  \node[above] at (1,0.1) {$S^2\times S^6$};
  \node[above] at (2,0.1)   {$S^2$};
\end{tikzpicture}
    \caption{$S^9$ as a fibration of $S^2\times S^6$ over the interval. At the extremes, either $S^2$ or $S^6$ shrink.}
    \label{fig:S9}
\end{figure}
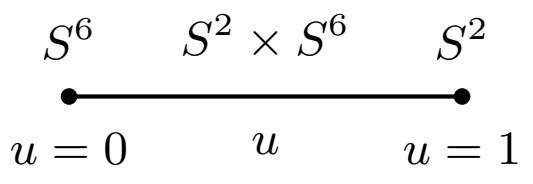

As mentioned in \autoref{sec:introduction}, the family of geometries of interest is asymptotically flat. Using the asymptotically Cartesian coordinates $x^M$ defined by the harmonic gauge condition \eqref{eq:harmonic_gauge}, one has
\begin{align}
    g_{MN}(x)=\delta_{MN}+o\left(1\right)
\end{align}
where the residual is given in small-$o$ notation with respect to the variable $1/r$, with $r=\sqrt{x^Mx_M}$. Asymptotically, $r=\text{const.}$ surfaces are topological $S^9$, as depicted in \autoref{fig:S9}. Let us now introduce the sources (which we will define here as the growing modes of the fields). We consider the following growing modes
\begin{align} \label{eq:leading}
\begin{aligned}
    y(x)&=y_0r^{-8}+o\left(r^{-8}\right),\\
    c(x)&=c_0r^3u^3+o\left(r^3\right),\\
    b(x)&=o\left(r^3\right),\\
    \tau(x)&=\tau_0r^2\left(1+2u^2\right)-\eta+o(1).
\end{aligned}
\end{align}
Notice that, while $y$ is decreasing, the dilaton $\phi=\log y$ grows in absolute value.
We left all residuals in small-$o$ notation, as we are agnostic on their fall-offs.
The constants $y_0,c_0$ and $\tau_0$ have a precise meaning on the matrix model side and they are all fixed among different saddles: they are given by combinations of $\mu$, the dimensionful parameter associated with the mass deformation, and of the dipole moment $P$ of the electrostatic configuration, fixed by the size $N$ of the matrices.
\begin{align} \label{eq:coeff}
    y_0=\frac{2^{14}P}{\pi^2\mu^7},\quad c_0=-\frac{\mu}{3},\quad \tau_0=\frac{\mu^2}{64}.
\end{align}
$\eta$ instead is a free parameter whose value can be freely changed through $SL\left(2,\mathds{R}\right)$ duality\footnote{A real form of $SL\left(2,\mathds{C}\right)$, as results from the complexification of the fields that always happens in Euclidean signature.} \cite{komatsu2025einsteingravitymatrixintegral}.

We would like now to study the remainders of \eqref{eq:leading} for an arbitrary solution to the equations of motion. The goal is to find the largest remainder that is not set to zero by those. The matter equations are
\begin{align}
 \label{eq:EOM1}\nabla^2\tau&=y|d\tau|^2+\tfrac12|G_3|^2,\\
 \label{eq:EOM2}\nabla^2y&=-\nabla_M(y^2\nabla^M\tau)-|H_3|^2+y\langle H_3,G_3\rangle,\\
 \label{eq:EOM3}-\text{i}\,d\widetilde F_7&=d*_g\left[ydc_2+\left(y\tau-1\right)H_3\right]=0,\\
 \label{eq:EOM4}d \widetilde H_7&=d*_g\left[\left(1-y\tau\right)dc_2+\tau\left(2-y\tau\right)H_3\right]=0.
\end{align}
where $\widetilde{F}_7$ and $\widetilde{H}_7$ are the electric seven-forms associated with $D1$- and $F1$-brane charges respectively (see \eqref{eq:stress_tensors}). To proceed, we neglect the metric fluctuations, showing later that those are subleading. 
\paragraph{Spherical harmonics}
In flat space, it is convenient to study the Laplacian in $S^9$ spherical harmonics; however, among all of those we have to consider only the $SO(3)\times SO(7)$ invariant ones thanks to the assumed symmetry. Concretely, the Laplacian associated with the flat metric
is given by
\begin{align}
    \nabla^2_0=\frac{1}{r^9}\partial_r\left(r^9\partial_r\right)+\frac{1}{r^2}\nabla_{S^9}^2
\end{align}
where, for scalar fields with no dependence on $S^2$ and $S^6$ angles, one has
\begin{align}
    \nabla_{S^9}^{2}\Big|_{SO(3)\times SO(7)}=\left(1-u^2\right)\partial_u^2
+\left(\frac{2}{u}-9u\right)\partial_u
\end{align}
The eigenfunctions of this operator are given in terms of Jacobi polynomials \cite{Koornwinder1973} $Y_{L}(u)\propto P^{\left(5/2,1/2\right)}_{L/2}\left(2u^2-1\right)$, $L=0,2,4,\dots$, normalized to unit leading coefficient. If we consider a fluctuation given by the mode $r^\lambda Y_L(u)$, its Laplacian is given by
\begin{align}
    \nabla_0^2\left[r^\lambda Y_L(u)\right]=\left[\lambda\left(\lambda+8\right)-L\left(L+8\right)\right]r^{\lambda-2}Y_L(u)
\end{align}
which is vanishing for $\lambda=L,-L-8$. As our boundary conditions rule out any growing mode for scalars, homogeneous modes can only appear for $\lambda=-L-8$. A similar analysis can be carried out for the flat space Laplacian of $SO(3)$ invariant two-forms, introducing the differential operator
\begin{align}
    *_gd*_gd\left[f(r,u)\omega_2\right]\equiv\mathcal{D}^2\left[f(r,u)\right]\omega_2
\end{align}
In flat space this reduces to
\begin{align} \label{eq:operator_twoforms}
\begin{aligned}
    \mathcal{D}^2_0&=\partial_r^2+\frac{5}{r}\partial_r+\frac{1}{r^2}\mathcal{D}_{S^9}^2,\\
    \mathcal{D}^2_{S^9}\Big|_{SO(3)\times SO(7)}&=\left(1-u^2\right)\partial_u^2-\left(5u+\frac{2}{u}\right)\partial_u
\end{aligned}
\end{align}
The eigenfunctions of $\mathcal{D}^2_{S^9}$ are again given in terms of Jacobi polynomials, $W_\ell\left(u\right)\propto u^3P_{\left(\ell-1\right)/2}^{(5/2,3/2)}\left(2u^2-1\right)$, $\ell=1,3,5,\dots$, with the same normalization. A two-form mode has eigenvalue
\begin{align}
\begin{aligned}
    &\mathcal{D}^2_0\left[r^\lambda W_\ell(u)\right]\\
    &\qquad=\left[\lambda\left(\lambda+4\right)-\left(\ell+2\right)\left(\ell+6\right)\right]r^{\lambda-2}W_\ell(u).
\end{aligned}
\end{align}
This vanishes for $\lambda=\ell+2,-\ell-6$. The leading spherical harmonics are
\begin{align}
\begin{aligned}
    &Y_0=1,\quad Y_2=u^2-\frac{3}{10},\quad\dots\\
    &W_1=u^3,\quad W_3=u^3\left(u^2-\frac{5}{12}\right),\quad\dots
\end{aligned}
\end{align}
The leading profiles of $y$ and $c$ solve the flat space equations $\nabla_0^2\left(y_0r^{-8}\right)=0$ and $\mathcal{D}_0^2\left(c_0r^3u^3\right)=0$, as expected.

The eigenvalue of the Laplacian allows us to distinguish sources and responses from forced modes. In particular, we use the following operational definitions: \emph{sources} and \emph{responses} are homogeneous modes, with a growing and decaying radial dependence respectively\footnote{Within the growing modes we include the constant ones too. The case of $y$ is different, indeed the leading term of the dilaton $\phi=\log y=-8\log r+\log y_0+o(1)$ is made of a logarithm and a constant term, which are non-decaying.}; non-homogeneous modes are instead \emph{forced}, either by other sources or other responses. We take the sources as fixed and equal to \eqref{eq:leading}, while the responses and the terms that they force are dependent on the saddle in consideration.

\paragraph{Analysis of the equations of motion} Let us start from \eqref{eq:EOM3}. In flat space we write it as
\begin{align}
\begin{aligned}
    0=\,&\left(y\tau-1\right)\mathcal{D}_0^2b\\
    &+y\mathcal{D}_0^2c+\langle dy,dc\rangle_0+\langle d\left(y\tau\right),db\rangle_0,
\end{aligned}
\end{align}
where the leading order of each term is: $y\mathcal{D}_0^2c=o\left(r^{-7}\right)$, $\mathcal{D}_0^2b=o\left(r\right)$, $y\tau\mathcal{D}_0^2b=o\left(r^{-5}\right)$, $\langle dy,dc\rangle_0=\mathcal{O}\left(r^{-7}\right)$ and $\langle d\left(y\tau\right),db\rangle_0=o\left(r^{-5}\right)$.
Therefore, the leading terms of the equations are driven by $\mathcal{D}_0^2b$. Let us suppose that the slowest mode of $b$ is $b(r,u)=b_0r^\lambda W_\ell(u)+o\left(r^\lambda\right)$, then the equation becomes
\begin{align}
\begin{aligned}
    &[\lambda\left(\lambda+4\right)-\left(\ell+2\right)\left(\ell+6\right)]b_0r^{\lambda-2}W_\ell\\&\quad=-24y_0c_0r^{-7}u^3+o\left(r^{-7}\right)+\mathcal{O}\left(r^{\lambda-8}\right)
\end{aligned}
\end{align}

\begin{widetext}
\begin{align} \label{eq:Einstein}
\begin{aligned}
    &R_{MN}=-\partial_{(M}y\,\partial_{N)}\tau-\frac{1}{2}y^2\partial_M\tau\,\partial_N\tau+\frac{1}{4}\left[2T_{MN}\left(H_3,G_3\right)-yT_{MN}\left(G_3,G_3\right)\right],\\
    &T_{MN}(A_3,B_3)=A_{PQ(M}\tensor{B}{_{N)}^{PQ}}-\frac{1}{12}g_{MN}A_{PQR}B^{PQR}
\end{aligned}
\end{align}
\end{widetext}

Now, we see that for $-5<\lambda\leq 2$, this equation has no solution. At $\lambda=-5$, the value of $b_0$ and $\ell$ are forced by the leading profiles of $y$ and $c$. In particular, $\ell=1$ and the eigenvalue of the Laplacian is $5-21=-16$. Thus, one finds that on a solution there is the forced mode
\begin{align}
    b=\frac{3}{2}c_0y_0r^{-5}W_1(u)+o\left(r^{-5}\right).
\end{align}

Notice that this mode is forced by a source and is thus independent of the saddle. 

We now have leading profiles for all matter fields, $b$ included, and we can evaluate the leading correction to the metric tensor induced through Einstein equations \eqref{eq:Einstein}. The stress-tensor has leading term of $\mathcal{O}\left(r^{-8}\right)$\footnote{Indeed, one has $\partial y\,\partial\tau=\mathcal{O}\left(r^{-8}\right)$, $y^2\left(\partial\tau\right)^2=\mathcal{O}\left(r^{-8}\right)$, $H_{MNP}=\mathcal{O}(r^{-8})$ as
\begin{align*}
    H_3=d\left(b_0 r^{-8}\frac{1}{2}\varepsilon_{ijk}x^idx^j\wedge dx^k\right)+o\left(r^{-8}\right),
\end{align*}
and $G_{MNP}=\mathcal{O}\left(1\right)$, led by $dc_2$
\begin{align*}
    G_3=3c_0\,dx^1\wedge dx^2\wedge dx^3+o(1).
\end{align*}
Then $T_{MN}\left(H_3,G_3\right)=\mathcal{O}\left(r^{-8}\right)$ and similarly $yT_{MN}\left(G_3,G_3\right)=\mathcal{O}\left(r^{-8}\right)$.}, hence
\begin{align} \label{eq:leading_R}
    R_{MN}=\mathcal{O}\left(r^{-8}\right).
\end{align}
Let us introduce a metric perturbation,
\begin{align}
    g_{MN}=\delta_{MN}+k_{MN},\qquad k_{MN}=o(1),
\end{align}
The harmonic gauge choice $\square_g x^M=0$ yields \cite{Carroll:2004st}
\begin{align}
    &R_{MN}=-\frac{1}{2}\nabla^2_0k_{MN}+o(k),\\\label{eq:harmonic_gauge_lin}
    &\partial^Mk_{MN}-\frac{1}{2}\partial_Nk_{PQ}\delta^{PQ}=o(k),
\end{align}
Similarly to above, we can decompose $k_{MN}$ into $SO(10)$ spherical harmonics $\mathcal{Y}_j\left(x\right),$ $j=0,1,2,\dots$ which, however, are not $SO(3)\times SO(7)$ invariant as, even though the metric is an invariant tensor, its Cartesian components are not invariant scalar functions. Suppose $k_{MN}=k^{(j)}_{MN}r^\lambda\mathcal{Y}_j(x)$ is the leading mode of $k_{MN}$, we have
\begin{align}
    \nabla_0^2\left[r^\lambda \mathcal{Y}_j\left(x\right)\right]=\left[\lambda\left(\lambda+8\right)-j\left(j+8\right)\right]r^{\lambda-2}\mathcal{Y}_j(x).
\end{align}
Plugging this back inside \eqref{eq:leading_R}, since there are no modes with $-8<\lambda<0$ annihilated by the Laplacian, the leading correction is forced to be
\begin{align}
    k_{MN}=\mathcal{O}\left(r^{-6}\right).
\end{align}
This fluctuation produces corrections to asymptotic equations of motion consisting of at most a relative factor of $\mathcal{O}\left(r^{-6}\right)$. For example, returning to \eqref{eq:EOM3} we found a leading term of the equation of order $\mathcal{O}\left(r^{-7}\right)$ which implies that the metric correction is of order $\mathcal{O}\left(r^{-13}\right)$, negligible for our purposes. One can similarly check that, for the orders we are considering, the metric correction can be neglected in the remaining equations of motion \eqref{eq:EOM1}, \eqref{eq:EOM2}, \eqref{eq:EOM4} as well.

Let us now proceed with \eqref{eq:EOM4}. Neglecting metric corrections, we can write it as
\begin{align}
\begin{aligned}
(1-y\tau)\mathcal{D}_0^2c=&
-\tau(2-y\tau)\mathcal D_{0}^2b+\langle d\left(y\tau\right),dc\rangle_0\\
&+\langle d\left(y\tau^2-2\tau\right),db\rangle_0
\end{aligned}
\end{align}
At leading order we find, $\mathcal{D}_0^2c=-2\tau\mathcal{D}_0^2b+\dots=\mathcal{O}\left(r^{-5}\right)$, implying that the order of the first correction is
\begin{align}
    c(r,u)=c_0r^3u^3+\mathcal{O}\left(r^{-3}\right).
\end{align}

Moving to the equation of motion \eqref{eq:EOM1}, we have that $y|d\tau|^2=\mathcal{O}\left(r^{-6}\right)$, $\left|G_3\right|_0^2-9c_0^2=\mathcal{O}\left(r^{-6}\right)$ and thus
\begin{align}
    \nabla_0^2\tau-\frac{9}{2}c_0^2=\mathcal{O}\left(r^{-6}\right).
\end{align}
Let $r^{\lambda}Y_L(u)$ be the leading order correction to $\tau$, we have
\begin{align}
    \tau=\tau_0r^2\left(\frac{8}{5}Y_0+2Y_2\right)-\eta+r^{\lambda}Y_L
\end{align}
the second term in the parenthesis and $\eta$ are annihilated by $\nabla_0^2$, while the first one reproduces $9c_0^2/2$ using \eqref{eq:coeff}. All in all, we find that the first subleading correction is
\begin{align}
    \tau=\tau_0r^2\left(1+2u^2\right)-\eta+\mathcal{O}\left(r^{-4}\right).
\end{align}

The final equation to study is \eqref{eq:EOM2} which will determine the order of the leading correction to the exponential of the dilaton $y$. As one can check, all forcing terms are of order $\mathcal{O}\left(r^{-16}\right)$
\begin{align}
    \nabla_0^2y=\mathcal{O}\left(r^{-16}\right).
\end{align}
This gives rise to three coefficients of homogeneous modes,
\begin{align}
\begin{aligned}
    y(r,u)=&\,y_0r^{-8}+y_2r^{-10}Y_2(u)+y_4r^{-12}Y_4(u)\\
    &+\mathcal{O}\left(r^{-14}\right).
\end{aligned}
\end{align}
Our boundary data only fixes $y_0$ leaving $y_2$ and $y_4$ free, determined by the saddle in consideration.

\paragraph{Terms forced by the responses} Response coefficients induce sub-leading terms in other fields. For example, we can look again at equation \eqref{eq:EOM3} and study the $r^{-7}$ modes in $b$. At this order there exists a homogeneous mode, left free by the asymptotic analysis, $b_2r^{-7}W_1\left(u\right)$; while there is also a term forced by $y_2$ through the coupling with the leading profile of $c$, 
\begin{align}
\begin{aligned}\mathcal{D}_0^2b=&\,-24y_0c_0r^{-7}W_1(u)-36c_0y_2\,r^{-9}W_3(u)\\
    &+\mathcal{O}\left(r^{-11}\right)
\end{aligned}
\end{align}
which is solved by
\begin{align}
\begin{aligned}
    b=&\frac{3}{2}c_0y_0r^{-5}W_1(u)+\left(\frac{3}{2}c_0y_2 W_3(u)+b_2W_1(u)\right)r^{-7}\\
    &+o\left(r^{-7}\right).
\end{aligned}
\end{align}

At these orders, no logarithm has appeared in any expansion. The power-series ansatz would fail only if there was a forcing term for a mode that is in the kernel of the differential operator. In that case, multiplying the mode by $\log r$ takes it outside the kernel allowing one to solve for the mode as a function of the source.

\paragraph{Fall-off conditions} To summarize, in the previous paragraph we found that any solution satisfies
\begin{align}
\label{eq:solutions_falloff}
\begin{aligned}
g_{MN}&=\delta_{MN}+\mathcal{O}\left(r^{-6}\right),\\
    y&=y_0r^{-8}+\mathcal{O}\left(r^{-10}\right),\\
    c&=c_0r^3u^3+\mathcal{O}\left(r^{-3}\right),\\
    b&=\frac{3}{2}c_0y_0 r^{-5}u^3+\mathcal{O}\left(r^{-7}\right),\\
    \tau&=\tau_0r^2\left(1+2u^2\right)-\eta+\mathcal{O}\left(r^{-4}\right).
\end{aligned}
\end{align}
with the first response coefficients appearing at $r^{-10}$ inside $y$ and $r^{-7}$ inside $b$.
As we will set up a variational principle with fluctuations having a non-compact support, we need to choose fall-off rates for off-shell configurations. This choice must clearly be done so as to include all saddles that share the given boundary data \eqref{eq:leading}. An economical choice would be to adopt the fall-offs of solutions \eqref{eq:solutions_falloff}; however, the counterterm that will be constructed works for a slightly larger class
\begin{align} \label{eq:fluct_falloffs}
\begin{aligned}
    \delta g_{MN}&=\mathcal{O}\left(r^{-6}\right),\quad
    \delta y=\mathcal{O}\left(r^{-10}\right),\\
    \delta c_{MN}&=\mathcal{O}\left(r^{-3}\right),\quad
\delta B_{MN}=\mathcal{O}\left(r^{-9}\right),\\
    \delta\tau&=\mathcal{O}\left(r^{-2}\right).
\end{aligned}
\end{align}
where we recall that the order of the two-forms in Cartesian components is related to that relative to the round unit $S^2$ volume form $\omega_2$ through a factor of $r^{-2}$ (within the $SO(3)$ ansatz, one would have $\delta c=\mathcal{O}\left(r^{-1}\right)$ and $\delta b=\mathcal{O}\left(r^{-7}\right)$).

%% file: Counterterm.tex
In this section, we start by computing the regularized boundary term of the on-shell action, supplemented by the Gibbons--Hawking--York (GHY) boundary term. We then supply it with a well-studied pure gravity counterterm for the leading $R^8$ divergence, caused by the GHY term, and put forward three candidate matter counterterms that could cancel the $R^2$ divergence. All three are intrinsic to the cutoff surface and with at most two derivatives acting on matter fields. Then, we move to the variational principle and show that, imposing the action to be stationary selects a unique linear combination of the three, thus removing the ambiguity. Finally, we discuss the validity of the counterterm as the cutoff surface is deformed and elaborate on its non-analyticity property. We show that there exists a rewriting of the counterterm that is fully analytic (even if the final result appears a bit cumbersome) and appears automatically in the dual frame of \cite{Skenderis_2000,Kanitscheider_2008}.

\paragraph{Regularized bare boundary action} We start by noticing that, as pointed out in \cite{komatsu2025einsteingravitymatrixintegral,Okuda_2008}, the on-shell action can be written as the integral of an exact differential form. Indeed, substituting the trace of equation \eqref{eq:Einstein} inside the IIB bulk action, one finds
\begin{align}
    I_{\text{IIB}}=\frac{1}{2\kappa^2}\int_{\mathcal{M}_R}\left(\frac{1}{2}H_3\wedge *G_3-\frac{y}{4}G_3\wedge *G_3\right).
\end{align}
We can recast this in terms of the electric seven-forms of \eqref{eq:stress_tensors}, which are closed by \eqref{eq:EOM3}, \eqref{eq:EOM4}
\begin{align} \label{eq:exact}
\begin{aligned}
    I_{\text{IIB}}&=\frac{1}{8\kappa^2}\int_{\mathcal{M}_R} \left(H_3\wedge \widetilde H_7 +\text{i}\,dc_2\wedge\widetilde F_7\right)\\
    &=\frac{1}{8\kappa^2}\int_{\mathcal{M}_R}d\left(B_2\wedge \widetilde H_7+\text{i}\,c_2\wedge \widetilde F_7\right)
\end{aligned}
\end{align}
We are concerned here with the asymptotic boundary contribution, which we write on the cutoff surface including the GHY term
\begin{align} \label{eq:bdy_bare}
\begin{aligned}
    I_{\text{bdy}}=&\frac{1}{8\kappa^2}\int_{\partial\mathcal{M}_R}*_h\Big[\tau(2-y\tau)\,\langle B_2,\iota_nH_3\rangle_h\\
    &+(1-y\tau)\big(\langle B_2,\iota_n dc_2\rangle_h+\langle c_2,\iota_nH_3\rangle_h\big)\\&-y\,\langle c_2,\iota_n dc_2\rangle_h-8K\Big].
\end{aligned}
\end{align}
where the unit normal vector to $\partial\mathcal{M}_R$ is chosen with outward orientation and it satisfies $*_g1=n^\flat\wedge *_h1$. It follows that\footnote{From the decomposition into tangent and normal components $B_q=n^\flat\wedge \iota_nB_q+i^*_{\partial\mathcal{M}_R}B_q$.} $i^*_{\partial\mathcal{M}_R}*_gA_p=*_h\iota_nA_p$ . $K$ is the trace of the second fundamental form $K_{MN}=\tensor{\left(\delta-n\,n\right)}{_M^P}\nabla_Pn_N$, $K=h^{MN}K_{MN}$.

Recall that the cutoff surface is defined by $x^Mx_M=R^2$, where $x^M$ are the harmonic coordinates used in the previous section. Splitting the metric as $g_{MN}=\delta_{MN}+k_{MN}$ where $k_{MN}=\mathcal{O}\left(r^{-6}\right)$ we have
\begin{align} \label{eq:induced_metric}
    h_{ab}=R^2\left[\mathring{h}_{ab}+k_{ab}\right]
\end{align}
where $\mathring{h}_{ab}$ is the round metric on the unit $S^9$. The unit normal vector has dual one-form given in coordinates as
\begin{align}
\begin{aligned}
    g_{MN}n^N&=\frac{\partial_M r}{\sqrt{\partial_Mr\,\partial^Mr}}\\
    &=\hat x_M\left(1+\frac{1}{2}k_{rr}+\mathcal{O}\left(k^2\right)\right)
\end{aligned}
\end{align}
where $\hat x_M=x_M/R$, lowered through the Euclidean flat metric. The order $k^2$ correction can be neglected in the limit $R\rightarrow\infty$, being of order $\mathcal{O}\left(R^{-12}\right)$, subleading with respect to $*_h1=\mathcal{O}\left(R^9\right)$. The corresponding vector is
\begin{align}
    n^M=\left(1+\frac{1}{2}k_{rr}\right)\hat x^M-k_{MN}\hat x^N +\mathcal{O}\left(k^2\right).
\end{align}
Then, we can expand the boundary integrand term by term. For example,
\begin{align}
\begin{aligned}
    K&=\frac{1}{\sqrt{g}}\partial_M\left(\sqrt{g}\,n^M\right)\\
    &=\frac{9}{r}+\frac{11}{2r}k_{rr}+\frac{1}{2}\partial_rk_{rr}-\frac{\tensor{k}{^N_N}}{r}+\mathcal{O}\left(k^2\right)
\end{aligned}
\end{align}
where we also used the harmonic gauge condition \eqref{eq:harmonic_gauge_lin}. Expanding the remaining terms in their $1/R$ series, as presented in the previous section, and integrating over the asymptotic sphere yields
\begin{align}
\begin{aligned}
    I_{\text{bdy}}^m&=\frac{\Omega_9}{2\kappa^2}\Big[-\frac{9}{20}c_0^2y_0R^2-\frac{3}{10}c_0b_2-\frac{21}{800}c_0^2y_2\Big]\\
    &\qquad+o(1),\\
    I_{\text{GHY}}&=\frac{\Omega_9}{2\kappa^2}\Big[-18R^8-\frac{81}{20}y_0c_0^2R^2-\frac{63}{10} c_0b_2\\
    &\qquad+\frac{63}{1600}c_0^2y_2\Big]+o(1).
\end{aligned}
\end{align}
As a remark, not all terms in \eqref{eq:bdy_bare} contribute. For example $\tau(2-y\tau)\,\langle B_2,\iota_nH_3\rangle_h=\mathcal{O}\left(r^{-13}\right)$. The metric contributions were determined by solving the linearized Einstein equations $R_{MN}=-\frac{1}{2}\nabla^2_0k_{MN}+o(k)$ for the orders $\mathcal{O}\left(r^{-6}\right)$ and $\mathcal{O}\left(r^{-8}\right)$ of $k_{MN}$. The sum gives
\begin{align}
\begin{aligned}
    &I_{\text{bdy}}=\\
    &\frac{\Omega_9}{2\kappa^2}\Big[-18R^8-\frac{9}{2}y_0c_0^2R^2-\frac{33}{5}c_0b_2+\frac{21}{1600}c_0^2y_2\Big]\\
    &\quad+o(1).
\end{aligned}
\end{align}
The divergent terms are dependent only on the sources with the responses entering only in the finite parts. 

\paragraph{Candidate counterterms} We now list a set of counterterms that one could add to the action in order to cancel the divergences. We specify that the counterterm we are going to construct is adapted to the cutoff surface we specified, namely $x^Mx_M=R^2$, but it will also work for surfaces that approach it asymptotically fast enough, as we will see. We start from the divergence proportional to $R^8$, coming from the GHY term. We expect to be able to cancel this using an intrinsic pure gravity counterterm; we adopt a candidate put forward in \cite{Lau_1999,Mann_1999}, further discussed in \cite{Mann_2006}\footnote{There it was shown to give results equivalent to other choices, for example the same on-shell action as background subtraction for a Schwarzschild black hole.}
\begin{align} \label{eq:grav_ct}
    \mathcal{S}_{\text{ct}}^{g}=\frac{1}{2\kappa^2}\int *_h\frac{18}{\scriptr},\qquad \frac{1}{\scriptr}\equiv\sqrt{\frac{R_h}{72}}
\end{align}
where $R_h$ is the scalar curvature of the induced metric. $\scriptr$ is at leading order given by the cutoff radius,
\begin{align} \label{eq:scriptr_exp}
    \frac{1}{\scriptr}=\frac{1}{R}\Big[1-\frac{\tensor{k}{_a^a}}{18}+\text{tot. der.}+\mathcal{O}\left(k^2\right)\Big]
\end{align}
where the total derivative can be dropped. One readily sees that its leading term successfully cancels the $R^8$ divergence. We will justify its usage beyond the cancellation of divergences when discussing the variational principle. Adding it shifts the boundary contribution to the on-shell action by
\begin{align}
\begin{aligned}
    &I_{\text{bdy}}+I^g_{\text{ct}}\\
    &=\frac{\Omega_9}{2\kappa^2}\left[\frac{9}{10}y_0c_0^2R^2+\frac{51}{40}c_0b_2-\frac{231}{6400}c_0^2y_2\right]\\
    &\quad+o(1)
\end{aligned}
\end{align}

We now turn to matter fields. Here we list more candidates that could absorb the quadratic divergence. We limit ourselves to candidates satisfying the following criteria
\begin{itemize}
    \item Intrinsic and gauge-invariant.
    \item Polynomial and with at most two derivatives on matter fields.
    \item With a factor $\scriptr^{n}$ where $n\leq 1$.
\end{itemize}
Outside these assumptions, one could find an arbitrarily large number of counterterms that can cancel the divergences. However, the stationarity of the action will ensure the uniqueness of the finite contribution to the on-shell action up to functions of sources only.

We find three candidates
\begin{align} \label{eq:candidates}
    \begin{aligned}
&\frac{1}{\Omega_9}\int*_h\frac{1}{\scriptr}y\tau
=\frac{9}{40}y_0c_0^2R^2+\frac{63}{6400}c_0^2y_2-\eta y_0,\\
&\frac{1}{\Omega_9}\int*_h\scriptr\,y\left|d_hc_2\right|^2_h=\frac{63}{10}y_0c_0^2R^2-\frac{63}{200}c_0^2y_2,\\
&\frac{1}{\Omega_9}\int*_h\scriptr\langle d_hB_2,d_hc_2\rangle_h=\frac{189}{20}y_0c_0^2R^2+\frac{63}{10}\,c_0b_2.
\end{aligned}
\end{align}
As we expect the counterterm to satisfy the same symmetries of the action, we demand it to be invariant under $\eta$-shifts. We therefore replace $\tau\mapsto \tau+\eta$ for the first candidate. As anticipated, the finite part is different in the responses showing that these candidates are not equivalent.
Considering a general linear combination \eqref{eq:general_ct},
cancellation of divergences only fixes the following constraint
\begin{align} \label{eq:finiteness_constraint}
    \alpha+28\beta+42\gamma=-4,
\end{align}
leaving two coefficients free. 

\paragraph{Variational principle} Let $\psi$ denote a bosonic field in $\Psi\equiv\{g_{MN},B_2,c_2,y,\tau\}$ and $\delta\psi$ its off-shell fluctuation with the fall-offs of \eqref{eq:fluct_falloffs}. The variation of the action splits into two components: a bulk and a boundary term. Schematically
\begin{align}
    \delta_\psi \mathcal{S}_{\text{IIB}}=\int_{\mathcal{M}_R} E_\psi[\Psi]\delta\psi+\int_{\partial\mathcal{M}_R}\Pi_\psi[\Psi]\delta\psi
\end{align}
where $E_\psi[\Psi]$ and $\Pi_\psi[\Psi]$ are local functionals of the fields. The boundary term is assumed to only depend on $\delta\psi$, not on normal derivatives. A variation proportional to the normal derivative of fluctuations can only be produced by the Einstein--Hilbert action, but it is canceled by the GHY action. 
\begin{widetext}
\begin{align} \label{eq:general_ct}
    \mathcal{S}_{\text{ct}}=\frac{1}{2\kappa^2}\int_{\partial\mathcal{M}_R}*_h\left[\frac{18}{\scriptr}+\alpha\frac{y}{\scriptr}\left(\tau+\eta\right)+\beta\scriptr y\left|d_hc_2\right|^2_h+\gamma\scriptr\langle d_hB_2, d_h c_2\rangle_h\right]
\end{align}
\end{widetext}

The idea is then to ensure that the counterterm cancels all terms of the variation that are not vanishing as the cutoff is removed $R\rightarrow\infty$. This ensures that satisfying the bulk equations of motion \eqref{eq:EOM1} to \eqref{eq:EOM4} is enough to make the action stationary.

Let us start by studying the variation of the action with respect to the metric. With the metric of the form \eqref{eq:induced_metric}, where $k_{ab}$ is replaced by a generic off-shell fluctuation $\delta k_{ab}$, we have the boundary terms
\begin{align}
\begin{aligned}
    &\delta_{g}\mathcal{S}_{\text{IIB}}+\delta_g \mathcal{S}_{\text{GHY}}\Big|_{\text{bdy}}\\
    &=\frac{1}{2\kappa^2}\int_{\partial\mathcal{M}_R}*_h\big(K^{ab}-Kh^{ab}\big)\delta h_{ab}\\
    &=-\frac{8}{2\kappa^2}\int_{S^9} d^9x\, \sqrt{\mathring{h}}\,R^8\tensor{\delta k}{_a^a}+o(1)
\end{aligned}
\end{align}
and one readily sees cancellation with the variation of the counterterm, from \eqref{eq:grav_ct} expanded as \eqref{eq:scriptr_exp}. Variation of the matter counterterm \eqref{eq:general_ct} with respect to the metric is $o(1)$, vanishing. This proves stationarity of the total action with respect to the variation of the metric
\begin{align}
\delta_g\mathcal{S}_{\text{bdy}}^{\text{ren}}=\delta_g\mathcal{S}_{\text{IIB}}^{\text{bdy}}+\delta_g\mathcal{S}_{\text{GHY}}+\delta_g\mathcal{S}_{\text{ct}}=0.
\end{align}

Consider now the variation with respect to matter fields. The boundary term is
\begin{align}
\begin{aligned}
    &\delta_{m}\mathcal{S}_{\rm IIB}\Big|_{\text{bdy}}\\
    &=\frac{1}{2\kappa^2}\int_{\partial\mathcal{M}_R}*_h\,\Big[-\partial_n\tau\delta y-\big(\partial_n y+y^2\partial_n\tau\big)\delta\tau
\\
&\qquad\qquad\qquad+\langle\delta B_2,\iota_n\left[(1-y\tau)G_3+\tau H_3\right]\rangle_h\\
&\qquad\qquad\qquad+\langle\delta c_2,\iota_n\left[(1-y\tau)H_3-ydc_2\right]\rangle_h\Big],
\end{aligned}
\end{align}
where the variations of $B_2$ and $c_2$ multiply the respective electric seven-forms $\widetilde H_7$ and $\widetilde F_7$. 

The terms with $\delta \tau$ and $\delta c_2$ are of order $\mathcal{O}\left(R^{-2}\right)$ and already
vanish as $R\rightarrow\infty$ without the need for a counterterm.

We are left with $\delta y$ and $\delta B_2$ terms only,
\begin{align} \label{eq:matter_variation}
\begin{aligned}
    &\delta_{m}\mathcal{S}_{\rm IIB}\Big|_{\text{bdy}}=\frac{1}{2\kappa^2}\int_{\partial\mathcal{M}_R}*_h1\,
\\&\Big[-\partial_n\tau\delta y+\langle\delta B_2,\iota_n\left[(1-y\tau)G_3+\tau H_3\right]\rangle_h\Big]
\end{aligned}
\end{align}
which are $\mathcal{O}\left(R^0\right)$, finite.
Thus, factors in front of fluctuations only contribute through their leading profiles. 

\begin{widetext} 
    \begin{align} \label{eq:final_counterterm}
        \mathcal{S}_{\text{ct}}=\frac{1}{2\kappa^2}\int_{\partial\mathcal{M}_R}*_h\left[\frac{1}{\scriptr}\left(18+2y\left(\tau+\eta\right)\right)-\frac{\scriptr}{7}\langle d_hB_2,d_hc_2\rangle_h\right],\qquad \frac{1}{\scriptr}=\sqrt{\frac{R_h}{72}}
    \end{align}
\end{widetext}

For the $\delta y$ factor,
\begin{align}
&\begin{aligned}
\partial_n\tau&=\partial_r\left(\tau_0r^2\left(1+2u^2\right)-\eta+\mathcal{O}\left(r^{-4}\right)\right)\\
    &=\frac{2}{R}\left(\tau+\eta\right)+\mathcal{O}\left(R^{-5}\right)\\
    &=\frac{2}{\scriptr}\left(\tau+\eta\right)+\mathcal{O}\left(R^{-5}\right)
\end{aligned}
\end{align}
which matches the form of the $\alpha$ counterterm in \eqref{eq:general_ct} with $\alpha=2$. Observe that one cannot similarly relate the leading profile of $\partial_n\tau$ to that of the $\beta$ counterterm, $\scriptr\left|d_h c_2\right|_h^2$, due to the different $u$ dependence of the leading profiles.

For the $\delta B_2$ factor we proceed similarly
\begin{align}
\begin{aligned}
\iota_n\big[(1-y\tau)&G_3+\tau H_3\big]
=\left(3c_0R^2u^3+\mathcal{O}(R^{-4})\right)\omega_2.
\end{aligned}
\end{align}
Consider the identity, valid for the round metric $h_0=R^2\mathring{h}$,
\begin{align}
    *_{h_0}d_{h}*_{h_0}d_h\left(u^3\omega_2\right)=\frac{1}{R^2}\mathcal{D}_{S^9}^2\left(u^3\right)\omega_2=-21\frac{u^3}{R^2}\omega_2
\end{align}
where $\mathcal{D}_{S^9}^2$ was defined in \autoref{sec:data}, \eqref{eq:operator_twoforms}. Hence, the result is
\begin{align}
\begin{aligned}
    \iota_n\big[(1-y\tau)G_3&+\tau H_3\big]\\
    &=-\frac{1}{7}\scriptr *_hd_h*_hd_hc_2+\mathcal{O}\left(R^{-4}\right)
\end{aligned}
\end{align}
Upon integration by parts, this reproduces the $\gamma$ counterterm with $\gamma=-1/7$. Therefore, we find that the linear combination \eqref{eq:general_ct} that makes the action stationary is
\begin{align}
\left(\alpha,\beta,\gamma\right)=\left(2,0,-1/7\right),
\end{align}
which satisfies the constraint required for finiteness \eqref{eq:finiteness_constraint}. The resulting counterterm is \eqref{eq:final_counterterm}.

\paragraph{Cutoff dependence} We now ask for which cutoff surfaces the counterterm \eqref{eq:final_counterterm} is valid. Up to now we checked explicitly the cancellation of divergences working with the cutoff $x^Mx_M=R^2$ with $x^M$ being harmonic asymptotically Cartesian coordinates. 

However, we also checked that the variation of the boundary term of the total action vanishes over any saddle and any fluctuation in the fall-off class \eqref{eq:fluct_falloffs}, $\delta\mathcal{S}_{\text{bdy}}=0$. Among such fluctuations, we have the variation of fields under the diffeomorphisms with infinitesimal vector field that satisfies
\begin{align}
    \xi^M=\mathcal{O}\left(r^{-5}\right).
\end{align}
Indeed, the variation of the metric is $\delta g_{MN}=2\nabla_{(M}\xi_{N)}=\mathcal{O}\left(r^{-6}\right)$, for scalar fields $\delta \psi=\xi^M\partial_M\psi=\mathcal{O}\left(r^{-6}\psi\right)$ and for two-forms $\delta A_{MN}=\mathcal{L}_\xi A_{MN}=\mathcal{O}\left(A_{MN}r^{-6}\right)$. All these variations are within our assumed fall-offs. 

The consequence is that one can choose any cutoff related to $x^Mx_M=R^2$ by a diffeomorphism with generating vector field $\xi^M=\mathcal{O}\left(r^{-5}\right)$. Once the gauge is changed from the harmonic one through one such diffeomorphism, the values of the on-shell action boundary term and that of the counterterm will change individually, but stationarity guarantees that their sum stays the same.

\paragraph{Analytic rewriting} As written, the counterterm \eqref{eq:final_counterterm} is not analytic in the induced metric $h_{ab}$. This comes as a consequence of the fact that the metric in asymptotically flat spacetimes has no asymptotic length scale: as the radius of the cutoff goes to infinity $R\rightarrow\infty$, the curvature of the sphere becomes vanishingly small $R_h\rightarrow 0$ allowing for no expansion of the square root. 

However, it turns out that there exists a rewriting of the same counterterm that is analytic. Essentially, the idea boils down to capturing the leading $R^{-2}$ decay through the dilaton,
\begin{align}
    \sqrt{R_h}=e^{\phi/8}\sqrt{e^{-\phi/4}R_h}.
\end{align}
The argument of the square root does not vanish anymore in the limit $R\rightarrow\infty$, allowing for an expansion of the square root. This combination appears automatically in the \emph{dual frame} variables \cite{Skenderis_2000,Kanitscheider_2008,Antoine}, defined via a Weyl transformation\footnote{Compared to the definition of \cite{Kanitscheider_2008} we used $y_0$ instead of $N$, the number of $D$-instantons. However, the latter is given by the integral of the electric axion nine-form (whose conservation is guaranteed by equation \eqref{eq:EOM2}) and is indeed proportional to $y_0$. Moreover, the term $+1/2$ added to $w$ is required in passing from Einstein to string frame, where the transformation of \cite{Kanitscheider_2008} is given.}
\begin{align}
    g_{d}=\left(e^\phi/y_0\right)^wg,\quad w=-\frac{2}{7-p}\Big|_{p=-1}+\frac{1}{2}=\frac{1}{4}
\end{align}
The leading metric profile is then transformed as follows
\begin{align}
\begin{aligned}
    g_d&=r^{-2}\left(1+\mathcal{O}\left(r^{-2}\right)\right)\left(dr^2+r^2d\Omega_9^2+\mathcal{O}\left(r^{-6}\right)\right)\\
    &=\frac{dr^2}{r^2}+d\Omega_9^2+\mathcal{O}\left(r^{-2}\right),
\end{aligned}
\end{align}
and it has the same form of the general $AdS_{p+2}\times S^{8-p}$ discussed in \cite{Kanitscheider_2008}, but with $p=-1$. It describes a spacetime asymptotic to $\mathds{R}\times S^9$, rather than to $\mathds{R}^{10}$: the $S^9$ radius does not blow up anymore as the cutoff is removed, $R\rightarrow\infty$. The scalar curvature in Einstein frame is related to that of the dual frame by
\begin{align}
    \begin{aligned}
    y_0^{1/4}e^{-\phi/4}R_h&=R_{h_d}+2\nabla^2_{h_d}\phi-\frac{7}{8}\langle d\phi,d\phi\rangle_{h_d}\\
    &\equiv 72\left(1+\epsilon\right)
\end{aligned}
\end{align}
where $\epsilon$ vanishes in the limit $r\rightarrow\infty$, indeed it is of order $\mathcal{O}\left(r^{-2}\right)$ as one can see from the left-hand side, recalling that $r^8y/y_0-1=\mathcal{O}\left(r^{-2}\right)$. We can thus expand the square root as
\begin{align}
    \sqrt{1+\epsilon}=1+\frac{\epsilon}{2}-\frac{\epsilon^2}{8}+\frac{\epsilon^3}{16}-\frac{5\epsilon^4}{128}+\mathcal{O}\left(r^{-10}\right)
\end{align}
which enters the full counterterm written in dual frame fields \eqref{eq:dual_frame_ct}, where we restored the dilaton $\phi$ through $y=e^{\phi}$ to underline the analyticity of the result. The order at which the expansion of the square root is truncated is decided based on the order of the multiplying factors. For example, a factor of $e^{-\phi}=\mathcal{O}\left(r^{8}\right)$ demands the full fourth order expansion. In the second line we plugged in the value of $\epsilon$ and rewrote the result in the Einstein frame, obtaining a fully analytic counterterm that makes the action stationary --- even if very unwieldy, with fine-tuned coefficients.
\begin{widetext}
    \begin{align} \label{eq:dual_frame_ct}
    \begin{aligned}
        \mathcal{S}_{\rm ct}&=\frac{1}{2\kappa^2}\int_{\partial\mathcal{M}_R}*_{h_d}\Bigg[y_0e^{-\phi}\Big(18+9\epsilon-\frac{9}{4}\epsilon^2+\frac{9}{8}\epsilon^3-\frac{45}{64}\epsilon^4\Big)\\
        &\qquad+2y_0\left(\tau+\eta\right)\Big(1+\frac{1}{2}\epsilon\Big)-\frac{1}{7}\,e^{-\phi/2}y_0^{1/2}\Big(1-\frac{1}{2}\epsilon\Big)\langle d_hB_2,d_hc_2\rangle_{h_d}\Bigg]\\
        &=\begin{aligned}[t]
\frac{1}{2\kappa^2}\int_{\partial\mathcal{M}_R}*_h\Bigg[\;&\frac{315}{64}\,\frac{e^{\phi/8}}{y_0^{1/8}}+\frac{35}{128}\,y_0^{1/8}e^{-\phi/8}R_h-\frac{35}{18432}\,y_0^{3/8}e^{-3\phi/8}R^2_h+\frac{7}{663552}\,y_0^{5/8}e^{-5\phi/8}R^3_h\\
&-\frac{5}{191102976}\,y_0^{7/8}e^{-7\phi/8}R^4_h+\frac{e^{9\phi/8}}{y_0^{1/8}}\left(\tau+\eta\right)+\frac{y_0^{1/8}}{72}\,e^{7\phi/8}\left(\tau+\eta\right) R_h\\
&-\frac{3}{14}\,y_0^{1/8}e^{-\phi/8}\langle d_hB_2,d_hc_2\rangle_h+\frac{y_0^{3/8}}{1008}\,e^{-3\phi/8}R_h\langle d_hB_2,d_hc_2\rangle_h\Bigg].
\end{aligned}
    \end{aligned}
    \end{align}
\end{widetext}

\paragraph{Another counterterm} A candidate counterterm was recently put forward in \cite{najafian2026compactholographypolarizedikkt},
\begin{align} \label{eq:new_ct}
\begin{aligned}
    &\mathcal{S}'_{\text{ct}}\\
    &=\frac{1}{2\kappa^2}\int_{\partial\mathcal M_R}*_h\left[\frac{18}{\scriptr}-\frac{4y\tau}{\scriptr}-\frac{11}{70}\scriptr\,\langle d_hy,d_h\tau\rangle_h\right]
\end{aligned}
\end{align}
and shown to make the action stationary under a restricted class of fluctuations, given by the variations of the multipole coefficients of the potential associated with the BPS saddles, see \autoref{sec:electrostatics}. We can check this counterterm against the more general variations with fall-offs \eqref{eq:fluct_falloffs}. In particular, let us consider the following off-shell fluctuations in the responses\footnote{Adding the $W_3$ harmonic induced in $b$ by $\delta y_2$ would not change the result, due to the orthogonality with $W_1$.},
\begin{align}
    \delta y=r^{-10}Y_2(u)\delta y_2,\quad\delta b=r^{-7}W_1(u)\delta b_2.
\end{align}
The variation of the type IIB action is (from \eqref{eq:matter_variation})
\begin{align}
\begin{aligned}
\delta\mathcal S_{\rm IIB}&=\frac{1}{2\kappa^2}\int_{S^9}*_{S^9}1\\
&\left[-2\tau_0\left(\frac{8}{5}Y_0+2Y_2\right)Y_2\delta y_2+3c_0\frac{W_1^2}{u^4}\delta b_2\right]\\
&=\frac{\Omega_9}{2\kappa^2}\left[-\frac{63}{3200}c_0^2\,\delta y_2+\frac{9}{10}c_0\,\delta b_2\right].
\end{aligned}
\end{align}
where we used $\tau_0=9c_0^2/64$. Instead, \eqref{eq:new_ct} varies only with respect to $\delta y_2$,
\begin{align}
\begin{aligned}
\delta\mathcal S'_{\rm ct}&=\frac{1}{2\kappa^2}\int_{S^9}*_{S^9}1\\
&\left[-4\tau_0\Big(\frac85Y_0+2Y_2\Big)-\frac{44}{7}\tau_0Y_2\right]Y_2\delta y_2\\
&=-\frac{\Omega_9}{2\kappa^2}\frac{9}{128}\,c_0^2\,\delta y_2 .
\end{aligned}
\end{align}
The variation of the total action is
\begin{align}
\begin{aligned}
    \delta\mathcal{S}_{\text{IIB}}+\delta\mathcal{S}_{\text{ct}}'=\frac{\Omega_9}{2\kappa^2}\left[-\frac{9}{100}c_0^2\delta y_2+\frac{9}{10}c_0\delta b_2\right]
\end{aligned}
\end{align}
and it vanishes only if the responses are constrained as
\begin{align} \label{eq:responses_rel}
    \delta b_2 \overset{!}{=}\frac{c_0}{10}\,\delta y_2.
\end{align}
As we will see in \autoref{sec:electrostatics}, this is indeed satisfied by the family of supersymmetric solutions of the electrostatic problem but it is violated for general fluctuations.

\paragraph{Uniqueness of the response-dependent finite part}
In this last paragraph, we show uniqueness of the response-dependent finite part of the renormalized action obtained through the counterterm \eqref{eq:final_counterterm}. In particular, we show that other counterterms (that could for example appear with factors $\scriptr^n$, $n\geq 2$) which also satisfy the constraints of the variational principle can shift the value of the on-shell action by finite terms that depend only on the sources, not on the responses.

All responses can be varied through properly engineered off-shell fluctuations (they all enter in the fall-offs of \eqref{eq:fluct_falloffs}). Let $\mathcal{S}''_{\text{ct}}$ be any other counterterm; its difference with ours can be expressed as a function of sources and responses
\begin{align}
\mathcal{S}_{\text{ct}}[\Psi]-\mathcal{S}_{\text{ct}}''[\Psi]=\mathcal{F}\left[y_0,c_0,y_2,b_2,\dots\right]
\end{align}
where the ellipsis denotes other subleading responses. Notice that this difference is finite for on-shell configurations, but could have a diverging part over an arbitrary configuration. Stationarity fixes the gradient of $\mathcal{F}$ with respect to any response to be vanishing over on-shell configurations,
\begin{align}
    \delta_{\text{responses}}\mathcal{F}\Big|_{\text{on-shell}}=0
\end{align}
this is because, again, off-shell fluctuations can change any response. The result is that any other counterterm making the variational principle well-defined under generic fluctuations gives automatically the same result as ours when evaluated on any saddle, up to source-only dependent terms\footnote{These source-dependent terms may in principle differ between connected components of the space of saddles at fixed sources.}.

Notice that we can use symmetries to restrict such source dependent additions. For example, the $\alpha$ counterterm (first line of \eqref{eq:candidates}) has such a piece proportional to $\eta$. However, demanding the counterterm to inherit $\eta$-independence from the type IIB action we subtracted it.

\begin{widetext}
\begin{align} \label{eq:electrostatics}
\begin{aligned}
&ds^2_\Sigma=-f\frac{V''}{\dot V}\,\big(d\rho^2+dz^2\big),\qquad
r_2^2=-f\frac{\rho V''\dot V}{\Delta},\qquad
r_6^2=3f\rho,\\
&y=-\mu^3\,\frac{3\dot V+\rho V''}{\rho\sqrt{-3\Delta\dot VV''}},\qquad
\tau=\frac{1}{\mu^3}\left[V'+\frac{3\rho\dot V\big(\dot V'^2+V''^2\big)}{3\dot V\dot V'+\sqrt{-3\Delta\dot VV''}}\right],\\
&b=-\frac{8}{3\mu}\left(z-\frac{\rho\dot V\dot V'}{\Delta}\right),\qquad
c=\frac{8}{3\mu^4}\left[V-\frac{\rho\dot V}{\Delta}\Big(V'\dot V'-3\dot VV''\Big)\right],
\end{aligned}\qquad
\begin{aligned}
f&\equiv\frac{8}{\mu^{5/2}}\left(\frac{\Delta\dot V}{-27V''}\right)^{1/4},\\[10pt]
\Delta&\equiv3\dot VV''+\rho\big(\dot V'^2+V''^2\big).
\end{aligned}
\end{align}
\end{widetext}

%% file: electrostatics.tex
In this section, we provide an application of the counterterm \eqref{eq:final_counterterm} to the computation of the on-shell action for the family of supersymmetric solutions dual to vacua of the polarized IKKT model. We start by briefly stating some of the properties of this family, referring to \cite{komatsu2025einsteingravitymatrixintegral}, then we move to the evaluation of its action as the sum of a bulk contribution and a renormalized boundary term.

As explained in \autoref{sec:introduction}, the geometry of the $1/2$-BPS saddles was derived in \cite{komatsu2025einsteingravitymatrixintegral} as an analytic continuation of the corresponding Lorentzian signature solution with $SO(5,2)\times SO(3)$ symmetry studied in \cite{D_Hoker_2016}. Recently, they were also obtained directly in Euclidean signature \cite{D_Hoker_2026,DHoker:2026eox}.

Using the ansatz \eqref{eq:ansatz}
and letting $(\rho,z)$ be coordinates for the base\footnote{With an abuse of notation, we use the same letters for the base coordinates, including $(r,u)$, as for the harmonic gauge ones introduced in \autoref{sec:introduction}. In fact, as we discuss they only differ by a diffeomorphism generated by $\xi^M=\mathcal{O}\left(r^{-5}\right)$ and therefore the difference drops out in the final result.} $\Sigma$, the solution has the explicit expression \eqref{eq:electrostatics},
depending only on a potential $V(\rho,z)$ which satisfies four-dimensional Laplace equation on $\Sigma\times S^2$ with $SO(3)$ symmetry, 
\begin{align} \label{eq:V_eq}
    V''+\ddot V+\frac{2}{\rho}\dot V=0 
\end{align}
where $'\equiv \partial_z$ and $\dot{}\equiv\partial_\rho$. The fibration over $\Sigma$ degenerates at $r_2^2=0$ and $r_6^2=0$. Avoiding a degenerate metric and smoothness force regularity constraints. Referring the reader to \cite{komatsu2025einsteingravitymatrixintegral} for details, as a result of the regularity analysis the $r_2^2=0$ curves are found at constant $z$, stretching between $\rho=0$ and an endpoint $\rho=\rho_s(z_s)$, 
while the $r_6^2=0$ one is at $\rho=0$. Moreover, the former must also satisfy $\dot V=0$ making them equipotential three-dimensional volumes in the electrostatic problem, referred to as \emph{disks}. The base space $\Sigma$ is then defined as the first quadrant $\rho,z\geq 0$, with $S^6$ collapsing at $\rho=0$ and $S^2$ collapsing at $z=0$ --- which is chosen as an infinite grounded disk $V|_{z=0}=0$ --- and at a set of disks $\{z_s\}_s$.
Fibering the $S^6$ over the geodesic going around each disk on the base gives a seven-cycle $\Sigma_7^s$, topologically $S^7$, which surrounds $D1$ and $F1$ branes. Moreover, fibering $S^2$ over an interval $z\in[z_s,z_{s+1}]$ between two consecutive disks yields a non-trivial three-cycle surrounding $D5$ and $NS5$ branes.

The solution to Laplace equation can be found through the method of images, extending $V(\rho,z)$ oddly across $z=0$. It can also be expanded in multipole moments: recalling the relation of $(r,u)$ with $(\rho,z)$, $z=ru$ and $\rho=r\sqrt{1-u^2}$, one has the equation
\begin{align}
\begin{aligned}
    &\left[\partial_r^2+\frac{3}{r}\,\partial_r+\frac{1}{r^2}\nabla^2_{S^3}\right]V=0,\\
    &\nabla^2_{S^3}\Big|_{SO(3)}=\left(1-u^2\right)\partial_u^2-3u\partial_u
\end{aligned}
\end{align}
where the Laplacian on $S^3$ has been restricted to $SO(3)$ invariant functions. Similarly to the spherical harmonics of \autoref{sec:data}, the solution is found in terms of Jacobi polynomials\footnote{We normalize them as Chebyshev polynomials of the second kind.} $U_\ell(u)\propto P^{(1/2,1/2)}_\ell(u)$, $\ell=1,3,\dots$ where the even values of $\ell$ have been excluded through parity under $z\rightarrow -z$. The first few polynomials are
\begin{align}
\begin{aligned}
    U_1&=2u,\quad U_3=8u^3-4u,\\
    U_5&=32u^5-32u^3+6u.
\end{aligned}
\end{align}
Then, for a mode $r^\lambda U_\ell(u)$ one has
\begin{align}
    \left[\lambda(\lambda+2)-\ell(\ell+2)\right]r^{\lambda-2}U_\ell(u)=0
\end{align}
with solutions $\lambda=\ell,-\ell-2$. The solution involves both an infinite series of multipoles and growing modes. The latter are chosen in order to guarantee positivity of metric and dilaton \cite{komatsu2025einsteingravitymatrixintegral},
\begin{align}
\begin{aligned}
V(r,u)=&-\frac{\eta\mu^3}{2}rU_1(u)-\frac{\mu^5}{2^9}r^3U_3(u)\\
&+\frac{1}{4\pi^2}\sum_{\ell=1,3,\dots}\frac{q_\ell}{r^{\ell+2}}U_\ell(u)
\end{aligned}
\end{align}
where $q_1\equiv P$ is the dipole moment, proportional to $y_0$ via \eqref{eq:coeff}. The expansion of the fields near $r\rightarrow\infty$ reveals corrections to the Euclidean metric that start at $\mathcal{O}\left(r^{-6}\right)$ \cite{komatsu2025einsteingravitymatrixintegral}. This shows that the gauge associated with these coordinates is related to harmonic gauge by a diffeomorphism with generating vector field $\xi^M=\mathcal{O}\left(r^{-5}\right)$, hence allowing us to apply the counterterm \eqref{eq:final_counterterm} with cutoff $r^2=R^2$. The expansions of $y$ and $b$ up to $q_3$ show that
\begin{align} \label{eq:responses_BPS}
    y_2 = 5\frac{2^{17}}{3\pi^2\mu^7}q_3,\qquad b_2=-\frac{2^{16}}{9\pi^2\mu^6}q_3.
\end{align}

\paragraph{Boundary contribution} We can now apply our counterterm and evaluate the boundary term as $R\rightarrow\infty$ for this class of solutions. Even if we work in the coordinate system of \cite{komatsu2025einsteingravitymatrixintegral}, rather than in harmonic gauge, we can apply the same formulas of the previous section because, as stated at the end of \autoref{sec:ct}, the stationarity of the action implies that the variation under the required diffeomorphism of the bare action result cancels with that of the counterterm. The result is
\begin{align} \label{eq:total_bdy}
\begin{aligned}
&I_{\text{bdy}}^{\text{ren}}=I_{\text{bdy}}+I_{\text{ct}}\\
    &=\frac{\Omega_9}{2\kappa^2}\left[-18R^8-\frac{9}{2}y_0c_0^2R^2-\frac{33}{5}c_0b_2+\frac{21}{1600}c_0^2y_2\right]\\
    &\,+\frac{\Omega_9}{2\kappa^2}\left[18R^8+\frac{9}{2}c_0^2y_0\,R^2+\frac{279}{40}c_0b_2-\frac{189}{6400}c_0^2y_2\right]\\
    &=\frac{\Omega_9}{2\kappa^2}\left[\frac{3}{8}c_0b_2-\frac{21}{1280}c_0^2y_2\right]\\
    &=\frac{1}{2\kappa^2}\frac{2^7\pi^3}{3\mu^5}q_3
\end{aligned}
\end{align}
where in the last line we substituted the responses \eqref{eq:responses_BPS} and $\Omega_9=\pi^5/12$.

\paragraph{Bulk contribution} We now turn to the evaluation of the bulk contribution, using the closed-form expression \eqref{eq:exact}
\begin{align}
I_{\text{IIB}}&=\frac{1}{8\kappa^2}\int_{\mathcal{M}_R}d\left(B_2\wedge \widetilde H_7+\text{i}\,c_2\wedge \widetilde F_7\right).
\end{align}
Let us start by noticing that the potentials $c_2$ and $B_2$ are not regular in the chosen gauge. This is because, at the disks they approach
\begin{align} \label{eq:bundle_data}
    c_s=\frac{8V_s}{3\mu^4},\quad
    b_s=-\frac{8z_s}{3\mu}
\end{align}
where $V_s$ is the value of the potential at the $s$ disk and $z_s$ is its height. In coordinates that are regular at the disks, the forms $B_2$ and $c_2$ blow up there. The issue is solved by covering the spacetime manifold with patches, $\mathcal{U}_j$ where $j=0,1,\dots s_f$ where $s_f$ is the number of conducting disks (with the grounded one $z=0$ excluded). The chart $\mathcal{U}_s$ with $s=1,\dots s_f$ surrounds the $s$-th disk and can be defined as a tubular neighborhood of $0\leq\rho\leq \rho_s,\,z=z_s$ on the base. $\mathcal{U}_0$ is instead defined as the complementary region and it is the only chart intersecting with the cutoff surface $\partial\mathcal{M}_R$. See \autoref{fig:charts}.
\begin{widetext}
\begin{align} \label{eq:stress_BPS}
\begin{aligned}
&\begin{aligned}
    -\text{i}\widetilde F_7=-\frac{2^9}{\mu^6}\Bigg[&15\rho^2\left(\dot Vdz-V'd\rho\right)+d\left(\rho^3\left(5V'+\frac{3\dot V\dot V'}{V''}\right)\right)\Bigg]\wedge\omega_6
\end{aligned}\\
&\begin{aligned}
    &\widetilde H_7=\frac{2^8}{\mu^9}\Bigg[30\rho^2\left(\dot V V'dz-\frac{1}{2}\left(V'^2-\dot V^2\right)d\rho\right)+d\left(\rho^3\left(5V'^2-3\dot V^2+\frac{6\dot V(\Delta+V'\dot V')}{V''}\right)\right)\Bigg]\wedge\omega_6
\end{aligned}
\end{aligned}
\end{align}
\end{widetext}
The gauge potentials are transformed between charts as
\begin{align}
    c^{(s)}=c^{(0)}-c_s,\quad b^{(s)}=b^{(0)}-b_s
\end{align}
so that each potential is regular in its domain. The bulk part of the on-shell action is then written as
\begin{align}
\begin{aligned}
    I_{\text{bulk}}&=\frac{1}{8\kappa^2}\sum_{0\leq i\leq s_f}\int_{\mathcal{U}_i}d\left(B_2^{(i)}\wedge\widetilde{H}_7+\text{i}\,c_2^{(i)}\wedge\widetilde F_7\right)\Bigg|_{\text{bulk}}\\
    &=\frac{1}{8\kappa^2}\sum_{1\leq s\leq s_f}\int_{\partial \mathcal{U}_s}\omega_2\wedge \left(b_s\widetilde H_7+\text{i}\,c_s\widetilde F_7\right)\\
    &=\frac{\Omega_2}{8\kappa^2}\sum_{1\leq s\leq s_f}\int_{\Sigma_7^s}\left(b_s\widetilde H_7+\text{i}\,c_s\widetilde F_7\right)
\end{aligned}
\end{align}
where $\Sigma_7^s$ represents the $7$-cycles given by the fibering of $S^6$ over the geodesic going around the $s$-th disk on the base, $\gamma_s$. In the last equality we used Poincaré duality on $\partial \mathcal{U}_s$ to recast the integral over $\Sigma_7^s$. On the family of BPS solutions, we can compute explicitly $\widetilde F_7$ and $\widetilde H_7$ from \eqref{eq:stress_tensors}, with the result being \eqref{eq:stress_BPS}. Dropping the exact part and integrating over $S^6$ one finds
\begin{figure}
    \centering
    \begin{tikzpicture}[scale=0.6]
  \fill[gray!25] (0,1.05) -- (2.6,1.05) arc (-90:90:0.55) -- (0,2.15) -- cycle;
  \fill[gray!25] (0,2.85) -- (1.725,2.85) arc (-90:90:0.55) -- (0,3.95) -- cycle;
  \draw[dashed] (0,1.05) -- (2.6,1.05) arc (-90:90:0.55) -- (0,2.15);
  \draw[dashed] (0,2.85) -- (1.725,2.85) arc (-90:90:0.55) -- (0,3.95);
  \node at (3.7,1.6) {$\mathcal{U}_1$};
  \node at (2.8,3.4) {$\mathcal{U}_2$};
  \node at (4.5,4.5) {$\mathcal{U}_0$};
  \draw[->] (0,0) -- (5.2,0) node[right] {$\rho$};
  \draw[->] (0,0) -- (0,4.9) node[above] {$z$};
  \draw[line width=1.6pt] (0,0) -- (0,4.6);
  \draw[line width=1.6pt] (0,0) -- (4.9,0);
  \node[left] at (-0.2,2.5) {$S^6\to 0$};
  \node[below] at (2.5,-0.2) {$S^2\to 0$};
  \node at (1,5) {$\vdots$};
  \draw[line width=1.6pt] (0,1.6) -- (2.6,1.6);
  \draw[line width=1.6pt] (0,3.4) -- (1.725,3.4);
\end{tikzpicture}
    \caption{Covering of the spacetime manifold in gauge patches $\cup_i \mathcal{U}_i=\mathcal{M}$ represented as tubular neighborhoods of disks on the base $\Sigma$. The dashed lines represent the boundary between a chart $\mathcal{U}_s$ and $\mathcal{U}_0$, projected onto the base.}
    \label{fig:charts}
\end{figure}
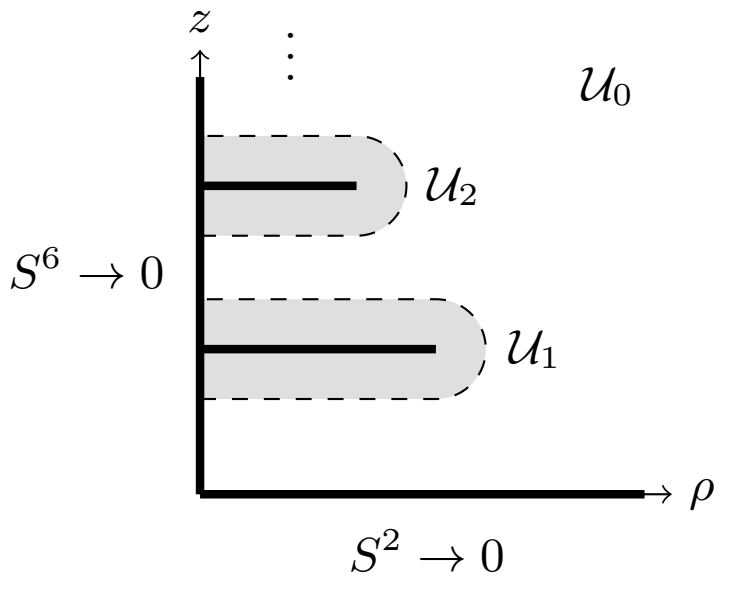
\begin{align} \label{eq:D1_charge}
\begin{aligned}
    \int_{\Sigma_7^s}\text{i}\widetilde F_7&=15\frac{2^9\,\Omega_6}{\mu^6}\int_{\gamma_s}V'\rho^2d\rho\\
    &=-15\frac{2^9\,\Omega_6}{\mu^6}\int_0^{\rho_s}\sigma_s(\rho)\rho^2 d\rho\\
    &=-\frac{2^{11}\pi^2}{\mu^6}Q_s.
\end{aligned}
\end{align}
In the second line we introduced the charge density of the conducting volume
\begin{align}
    \sigma_s(\rho)=\frac{1}{\Omega_2\rho^2}\frac{dQ_s}{d\rho}\equiv \lim_{z\rightarrow z_s^-}V'-\lim_{z\rightarrow z_s^+}V'.
\end{align}
It turns out that regularity fixes the value of $\rho_s(z_s)$ via the condition $\sigma_s(\rho_s)=0$ \cite{komatsu2025einsteingravitymatrixintegral}.

Let us turn to the flux of $H_3$, 
\begin{align}
\begin{aligned}
    \int_{\Sigma_7^s}\widetilde H_7&=15\frac{2^8\Omega_6}{\mu^9}\int_{\gamma_s}\left(V'^2-\dot V^2\right)\rho^2d\rho\\
    &=15\frac{2^8\Omega_6}{\mu^9}\int_{\gamma_s}V'^2\rho^2d\rho\\
    &=-15\frac{2^9\Omega_6}{\mu^9}\int_0^{\rho_s} V'_{\hat s}\sigma_s\rho^2d\rho\\
    &=\frac{2^{11}\pi^2}{\mu^9}F_s
\end{aligned}
\end{align}
where we dropped the $\dot V^2$ term as it is continuous across the disk, we introduced $V_{\hat s}'\equiv \frac{1}{2}\left(V'|_{z\rightarrow z^+_s}+V'|_{z\rightarrow z^-_s}\right)$ (which can be interpreted as minus the electric field with the contribution of the conducting volume excluded) and $F_s$ (the electric force on the disk). Plugging everything together we find
\begin{align} \label{eq:total_bulk}
\begin{aligned}
    I_{\text{bulk}}&=-\frac{1}{2\kappa^2}\frac{2^{14}\pi^3}{3\mu^{10}}\sum_s\left(Q_sV_s+z_s F_s\right)\\
    &=-\frac{1}{2\kappa^2}\frac{2^{14}\pi^3}{3\mu^{10}}\left[2\sum_sQ_s V_s\Big|_{\eta=0}+\frac{\mu^5q_3}{2^6}\right].
\end{aligned}
\end{align}
In the last step we used an identity for $\sum_sz_sF_s$ that we prove in \autoref{app:electro}. Let us observe that $I_{\text{bulk}}$ is $\eta$-independent. This is expected since, again, a shift by $\eta$ acts as a duality transformation and the integrand form \eqref{eq:exact} is pointwise invariant.

\paragraph{On-shell action} Summing the bulk contribution \eqref{eq:total_bulk} with the renormalized boundary one \eqref{eq:total_bdy}, one finds
\begin{align}
\begin{aligned}
    &I=-\frac{1}{2\kappa^2}\frac{2^{15}\pi^3}{3\mu^{10}}\sum_s Q_s V_s\Big|_{\eta=0}-\frac{1}{2\kappa^2}\frac{2^7\pi^3}{3\mu^5}q_3\\
    &=-\frac{\mathcal{K}}{2}\left(\sum_s Q_sV_s\Big|_{\eta=0}+\frac{\mu^5}{2^8}\,q_3\right),
\end{aligned}
\end{align}
where we introduced the constant $\mathcal{K}$ in order to ease the notation
\begin{align} \label{eq:K}
\mathcal{K}\equiv\frac{1}{2\kappa^2}\frac{2^{16}\pi^3}{3\mu^{10}}=\frac{2^9}{3\pi^4g_s^2\alpha'^4\mu^{10}}.
\end{align}
The final result matches the one expected on the matrix model side only up to the sign of the first term \cite[eq. 5.49]{komatsu2025einsteingravitymatrixintegral2}. We address this sign issue in the next section; this is the analog of the mechanism discussed in \cite{Hawking_1995} for electrically charged black holes.

%% file: canonical.tex
From a semi-classical point of view, the on-shell action appears formally as a saddle of the supergravity path integral
\begin{align}
    Z=\int_{\text{\eqref{eq:leading},\eqref{eq:fluct_falloffs}}} \mathcal{D}\Psi\,e^{-\mathcal{S}_{\text{IIB}}[\Psi]-\mathcal{S}_{\text{ct}}[\Psi]}
\end{align}
where \eqref{eq:leading} and \eqref{eq:fluct_falloffs} specify the boundary data and the fall-offs of off-shell configurations that we assumed throughout this note. $\Psi=\{g_{MN},B_2,C_2,\phi,\chi,\dots\}$ is the set of dynamical fields in the truncated sector $F_5=0$ that we are considering\footnote{\label{foot:other} The ellipsis includes $C_4$, the fermions of type IIB supergravity, as well as gauge fixing and Faddeev–Popov ghosts. However, none of these affect the classical analysis of this note. Fermions vanish on the saddles we are considering and enter the action at least quadratically. Similarly, $F_5$ enters quadratically and $C_4$ through the CS term $H_3\wedge F_3\wedge C_4$, but $H_3\wedge F_3=0$ on the ansatz \eqref{eq:ansatz}. Therefore, they enter neither the on-shell action nor its first variation around the saddles, whatever their fall-offs. Consequently, we do not have to specify their fall-offs for our argument, but they become relevant for a check of supersymmetry and for quantum corrections. With similar arguments, the ghosts and gauge fixing sectors can be shown to be important only when extending the analysis beyond the classical level.}.  The addition of the counterterm assumes first a regularization through a cutoff surface $\partial\mathcal{M}_R$, followed by the limit $R\rightarrow\infty$.

Now, observe that the path integral over a field $\psi\in\Psi$ naturally splits as a sum over bundles combined with a path integral over the continuous fluctuations. Let us refer to the bundle data here as chemical potentials and denote it as $\{\nu_\alpha\}_\alpha$ where $\alpha$ is an index for the potentials $C_2, B_2$. The supergravity path integral is then rewritten as a sum over grand-canonical partition functions
\begin{align}
\begin{aligned}
    Z&=\sum_{\{\nu_\alpha\}}\int_{\eqref{eq:leading},\eqref{eq:fluct_falloffs}} \mathcal{D}\Psi_{\{\nu_\alpha\}}\,e^{-\mathcal{S}_{\text{IIB}}-\mathcal{S}_{\text{ct}}}\\
    &=\sum_{\{\nu_\alpha\}}Z_{\text{gc}}[\{\nu_\alpha\}]
\end{aligned}
\end{align}
where $\Psi_{\{\nu_\alpha\}}$ denotes the remaining fields and the integral runs over continuous fluctuations with fixed transition functions. The bundle data is given by the periods of $dC_2$ and $dB_2$ over the non-trivial $3$-cycles and it is presented in \eqref{eq:bundle_data},
\begin{align}
    \left\{\nu_\alpha\right\}=\{c_s,b_s\}\sim\{V_s,z_s\}.
\end{align}
Each grand-canonical partition function sums over fluctuations with specified disk potentials $V_s$ and heights $z_s$, i.e. number of $D5$ and $NS5$ branes \cite{komatsu2025einsteingravitymatrixintegral}. While on the matrix model side $N_{NS5}^s$ is mapped to $N_s$ --- which is the dimension of the $s$-th $SU(2)$ irreducible representation entering the VEV of the $X^i\in \mathfrak{u}(N)$ matrix model fields at a supersymmetric vacuum ---, the number of $D5$ branes is left free in the choice of supersymmetric vacuum, which instead requires fixing the number of $D1$ branes,
\begin{align}
    n_s\sim N_{D1}^s\sim \int_{\Sigma_7^s}\widetilde F_7\sim Q_s
\end{align}
where $n_s$ is the multiplicity of the $s$-th representation entering the VEV of $X^i$. In the supersymmetric localization calculation, $n_s$ is fixed by introducing Lagrange multipliers for the eigenvalue densities \cite[eq. 5.23]{komatsu2025einsteingravitymatrixintegral2}; we can impose the analog constraint in the supergravity path integral through dualization of the RR potential $C_2$.

Dualization can be performed along the same lines as electromagnetic duality in four-dimensional Maxwell theory \cite{witten1995sdualityabeliangaugetheory,dijkgraaf1997leshoucheslecturesfields}, namely by changing variable to $f_3=dC_2$ and introducing a Lagrange multiplier six-form $\widetilde C_6$ that ensures the quantization of the $f_3$ periods. The path integral becomes
\begin{align} \label{eq:dualized_pathintegral}
\begin{aligned}
    Z=&\int_{\eqref{eq:leading},\eqref{eq:fluct_falloffs}}\Bigg[\prod_{\substack{\psi\in\Psi\\
    \psi\neq C_2}}\mathcal{D}\psi\Bigg]\mathcal{D}\widetilde C_6\,\mathcal{D}f_3\\
    &e^{-\mathcal{S}_{\text{IIB}}-\mathcal{S}_{\text{ct}}+\frac{\text{i}}{2\kappa^2}\left(\int_{\mathcal{M}_R} d\widetilde C_6\wedge f_3-\int_{\partial\mathcal{M}_R}\widetilde C_6\wedge f_3\right)},
\end{aligned}
\end{align}
the topological BF-coupling added to the action ensures both that $df_3=0$ and that its periods are quantized, as we will check. The new surface term is an addition to the counterterm that is required to preserve the stationarity of the action on-shell. The remaining terms in the truncated action, $\mathcal{S}_{\text{IIB}}+\mathcal{S}_{\text{ct}}$ (including the fermionic one) are all dependent on $C_2$ only through $dC_2$, which we replace with $f_3$. As $f_3$ appears quadratically, it can be integrated out yielding the relation
\begin{align} \label{eq:duality}
    d\widetilde C_6=\text{i}\widetilde F_7
\end{align}
from which $\widetilde C_6$ inherits a leading behavior and fall-off of fluctuations from \eqref{eq:fluct_falloffs}. In particular, writing $\widetilde C_6=\widetilde c\,\omega_6$, one finds $\delta \widetilde c = \mathcal{O}\left(r^{-3}\right)$.
Let us check the $\widetilde C_6$ equation of motion,
\begin{align}
\begin{aligned}
    &\int_{\mathcal{M}_R}d\delta\widetilde C_6\wedge f_3\\
    &\qquad=-\int_{\mathcal{M}_R}\delta\widetilde C_6\wedge df_3+\int_{\partial\mathcal{M}_R}\delta\widetilde C_6\wedge f_3
\end{aligned}
\end{align}
as on-shell $f_3$ is asymptotically order $f_3\sim r^3u^2 du\wedge \omega_2$ and the fall-off of $\delta\widetilde c$ is $\mathcal{O}\left(r^{-3}\right)$, the asymptotic boundary term contribution to the variation does not vanish and it would spoil stationarity. Therefore, we subtracted it through the counterterm
\begin{align}
    -\int_{\partial\mathcal{M}_R}\widetilde C_6 \wedge f_3
\end{align}
which is well-defined since $\widetilde C_6$ is globally-defined on the asymptotic $S^9$ (it has no non-trivial seven-cycles). Also, it is gauge invariant as only the leading profile of $f_3$, which is fixed to be a closed form by the boundary data, gives a contribution (assuming gauge transformations of $\widetilde C_6$ respect its fall-off). Now let us check the quantization of periods that comes with summing over the fluxes of $\widetilde C_6$. Let $\omega_{7\,s}\in H^7\left(\mathcal{M},\mathds{Z}\right)$ be a basis of harmonic forms; we split a general profile of $d\widetilde{C}_6$ as
\begin{align}
    d\widetilde{C}_6=\sum_{s}n_s\omega_{7\,s}+d\delta\widetilde C_6,\quad n_s\in\mathds{Z}
\end{align}
where $\delta\widetilde C_6$ is globally-defined. Then, the path integral includes a sum over the $\widetilde C_6$ bundles, i.e. over the values of $n_s$
\begin{align}
    Z=\text{\eqref{eq:dualized_pathintegral}}_{\widetilde C_6\rightarrow \delta\widetilde{C}_6}\sum_{\{n_s\}}e^{\frac{\text{i}}{2\kappa^2}\sum_s n_s\int\omega_{7\,s}\wedge f_3}
\end{align}
where the common factor represents the path integral \eqref{eq:dualized_pathintegral} where $\widetilde C_6$ is only integrated over its continuous component $\delta\widetilde C_6$. Summing over $n_s$ and using Poincaré duality\footnote{Actually, the manifold $\mathcal{M}_R$ has a boundary $\partial\mathcal{M}_R$, so the correct duality to use would be Poincaré--Lefschetz duality.} yields $\delta$-functions contributions enforcing\footnote{This is a formal quantization condition since the periods of $f_3$ (and of $\widetilde F_7$) are actually imaginary due to the choice of integration contour in the Euclidean path integral.}
\begin{align}
    \frac{1}{2\kappa^2}(2\pi)^6\alpha'^3g_s\int_{\Sigma_{3}^s}f_3\in 2\pi\mathds{Z}
\end{align}
where the factor comes from the quantization unit of $d\widetilde C_6=\text{i}\widetilde{F}_7$ periods. Substituting $2\kappa^2=\left(2\pi\right)^7g_s^2\alpha'^4$, one finds
\begin{align}
    \int_{\Sigma_{3}^s}f_3\in \left(2\pi\right)^2\alpha'g_s\mathds{Z}
\end{align}
which is the correct quantization condition for $dC_2$ \cite{komatsu2025einsteingravitymatrixintegral}. On more general grounds, the quantization of fluxes yields the correct units by the Dirac--Nepomechie--Teitelboim quantization condition \cite{Polchinski_1995}.

Finally, we can split the new representation of the formal supergravity path integral as a sum over the topological sectors of $\widetilde C_6$, which fix the charges $n_s\sim Q_s$. The result is a set of canonical partition functions, where the charges labeling each matrix model vacuum are fixed
\begin{align}
    Z=\sum_{\{z_s,Q_s\}}Z_{\text{can}}[\{z_s,Q_s\}].
\end{align}
Now, once the charges $\{z_s,Q_s\}$ are specified, we can compute the saddle contribution to the canonical partition function from the on-shell action. In particular, one has simply to compute the on-shell value of the BF-term
\begin{align}
\begin{aligned}
    &I_{\text{can}}\left[\left\{z_s,Q_s\right\}\right]=I\left[\{z_s,V_s\}\right]\\
    &\quad-\frac{\text{i}}{2\kappa^2}\left(\int_{\mathcal{M}_R} \text{i}\,\widetilde F_7\wedge dC_2+\int_{\partial\mathcal{M}_R} \text{i} \,\widetilde F_7\wedge C_2 \right)
\end{aligned}
\end{align}
which can be written in terms of bundle transition data using the covering depicted in \autoref{fig:charts}. In particular,
\begin{align}
    \begin{aligned}
        I_{\text{can}}&=I-\frac{1}{2\kappa^2}\sum_{1\leq s\leq s_f}\int_{\partial \mathcal{U}_s} c_s\omega_2\wedge \text{i}\widetilde F_7&\\
        &=I-\frac{\Omega_2}{2\kappa^2}\sum_{1\leq s\leq s_f}c_s\int_{\Sigma_7^s}\text{i}\widetilde F_7
    \end{aligned}
\end{align}
which has precisely the form of a Legendre transform from the $D1$ grand-canonical to the $D1$ canonical sector, as for the fixed-charge ensemble of electric black holes \cite{Hawking_1995}. We show in \autoref{app:electro} that the same result can be achieved with the Legendre transform of the on-shell action that swaps $V_s$ for their conjugate by means of identities from electrostatics. This derivation instead highlights how the fixed charge ensemble is achieved inside the formal supergravity path integral, working directly with the ten-dimensional type IIB supergravity fields.

We can end the calculation by plugging in \eqref{eq:D1_charge} and \eqref{eq:bundle_data}, obtaining
\begin{align} \label{eq:final}
\begin{aligned}
    &I_{\text{can}}=I+\mathcal{K}\sum_s Q_sV_s\\
    &=\frac{\mathcal{K}}{2}\left(\sum_sQ_sV_s\Big|_{\eta=0}-\frac{\mu^5}{2^8}q_3\right)\\
    &\qquad-\frac{2\pi\eta  N}{g_s}.
\end{aligned}
\end{align}
where $N$ is the number of $D$-instantons, given by 
\begin{align}
    N=\frac{2^7}{3\pi^5\mu^7\alpha'^4 g_s}P,\quad P=2\sum_s z_s Q_s
\end{align}
Equation \eqref{eq:final} matches the matrix model result, $\eta$-shift included \cite{komatsu2025einsteingravitymatrixintegral}, \cite[eq. 5.49]{komatsu2025einsteingravitymatrixintegral2}. It is worth noticing that the only $\eta$-dependence comes from the Legendre transform, while the grand-canonical on-shell action is $\eta$-invariant.

%% file: conclusion.tex
We constructed a counterterm for type IIB supergravity with symmetries and boundary conditions of the gravity dual to supersymmetric vacua of the polarized IKKT matrix model. To arrive at it, we started in \autoref{sec:data} by studying the asymptotic expansions of a general solution distinguishing the boundary data and the forced modes from the responses. We then stated a class of off-shell configurations given by fall-off rates \eqref{eq:fluct_falloffs} that are slightly slower than those of saddles and, in \autoref{sec:ct}, we constructed a counterterm by demanding that the variational principle be well-defined. This means ensuring the cancellation of the boundary contribution to the variation of the action under generic fluctuations (within the stated fall-off rates). We have also shown the existence of a rewriting of the counterterm that makes it fully analytic and discussed its uniqueness up to source-dependent finite terms. In \autoref{sec:electrostatics} we specialized to the supersymmetric saddles of interest for the polarized IKKT model and evaluated their renormalized on-shell action. Finally, in \autoref{sec:canonical} we introduced the dualization of the RR potential as a formal manipulation of the supergravity path integral that allowed us to expand it into fixed-charge sectors and we computed the saddle free-energy of a given sector. The result matches the polarized IKKT prediction for the on-shell action obtained in \cite{komatsu2025einsteingravitymatrixintegral2} through supersymmetric localization.

One possible (much harder) next step could consist of extending the analysis beyond the classical level, by comparing the one-loop determinant on both sides of the correspondence. Also, while stationarity is a necessary condition for the counterterm, one should also ensure that it is compatible with supersymmetry \cite{Freedman_2017,Genolini_2017}. Ensuring supersymmetry would allow one to further constrain the dependence on sources of the finite part of the on-shell action, left free by the uniqueness argument of \autoref{sec:ct}. For example, in \cite{Genolini_2017} it is shown that a supersymmetric Ward identity requires the addition of a new finite term for $\mathcal{N}=2$ gauged supergravity in five dimensions. Source-dependent terms can be fixed also in other ways, for example in \cite{dei2025onshellgravityactionlinear} the freedom left by the variational principle is fixed by demanding the on-shell action to reduce smoothly to the asymptotically $AdS_3$ result in the limit in which the linear-dilaton region is pushed to infinity ($R_\phi\rightarrow\infty$).  Finally, as the counterterm has been constructed from a general analysis of the asymptotics, it could be applied as well for studying non-supersymmetric saddles with the same sources and $SO(3)\times SO(7)$ symmetry (where \eqref{eq:responses_rel} need not hold). An example could be the linearized backreaction of the probe $D1$-brane configurations of \cite{hartnoll2024polarisedikktmatrixmodel}.
 We leave these questions for future work.

%% file: electro_app.tex
\section{Electrostatics Identities}
\label{app:electro}

In this appendix, we derive some identities useful in the analysis of the electrostatic problem associated with the supersymmetric saddles of \autoref{sec:electrostatics}. 

\begin{widetext}
\begin{align}\label{eqn:electrostatic-energy}
&\begin{aligned}
    \mathcal{E}[\{\sigma_s\};\{z_s\}]\equiv \frac{1}{2}\sum_{s_1,s_2}\int\int d\rho_1\,d \rho_2\,\rho_1\,\rho_2\,\sigma_{s_1}(\rho_1)\,\sigma_{s_2}(\rho_2)\,G_{2,\mathrm{im}}\left(\rho_1,\rho_2,z_{s_1},z_{s_2}\right)+\sum_s\Omega_2\int d \rho\,\rho^2\,\sigma_s(\rho)\, V_{\text{bg}}(z_s,\rho)
\end{aligned}\\
&\begin{aligned}
    V_{\text{bg}}=-\frac{\eta\mu^3}{2}rU_1(u)-\frac{\mu^5}{2^9}r^3U_3(u)\equiv V_{\text{bg},1}+V_{\text{bg},3}
\end{aligned}
\end{align}
\end{widetext}

\paragraph{Identity for $\sum_s z_s F_s$} Let us start by introducing the total energy of the electrostatic configuration in the half-space $z>0$, given by the binding energy of the conductors, including the contribution of the images in the grounded plane at $z=0$, together with their potential energy in the background, \eqref{eqn:electrostatic-energy}. The sums run only over physical conductors and the radial integrals over their supports. The Green function including the image charges is \cite[app. C]{komatsu2025einsteingravitymatrixintegral}
\begin{align}
    \begin{aligned}
    &G_{2,\mathrm{im}}(\rho_1,\rho_2,z_1,z_2)=\begin{aligned}[t]
    &G_2(\rho_1,\rho_2,z_1,z_2)\\
    &\quad-G_2(\rho_1,\rho_2,z_1,-z_2),
    \end{aligned}\\
    &G_2\left(\rho_1,\rho_2,z_{1},z_{2}\right)=\log\left(1+\frac{4\rho_1\rho_2}{\left(\rho_1-\rho_2\right)^2+(z_1-z_2)^2}\right).
    \end{aligned}
\end{align}
Here $G_2$ is obtained by angular integration of the four dimensional Green function. The subtracted term is the potential of the opposite image charge at $-z_s$ and enforces $V=0$ on the grounded plane.

We first study the energy under a uniform rescaling of the $(\rho,z)$ plane at fixed charges, with\footnote{The rescaled distributions in general do not give rise to equipotential disks away from $\lambda=1$. However, the energy variation at linear order caused by redistributing the charges within each disk is proportional to the potential differences at zeroth order (which vanish on each conductor).} $\sigma_{s,\lambda}(\rho)=\lambda^{-3}\sigma_s(\rho/\lambda)$. 
Denoting the binding-energy of the conductors system by $U$ and the background term by $W_{\text{ext}}\equiv W_{\text{ext},1}+W_{\text{ext},3}$, one has
\begin{align}
\begin{aligned}
    \mathcal{E}[\{\sigma_{s,\lambda}\};\{\lambda z_s\}]&=\lambda^{-2}U+\lambda^1W_{\text{ext},1}+\lambda^3W_{\text{ext},3}\\
    \lambda\frac{d\mathcal{E}}{d\lambda}\Big|_{\lambda=1}&=-2U+W_{\text{ext},1}+3W_{\text{ext},3}
\end{aligned}
\end{align}
The $\lambda$ factors come from the measure transforming as $d\rho_i\,\rho_i\rightarrow \lambda^2\,d\rho_i\,\rho_i$, the charge density as $\sigma_s\rightarrow\lambda^{-3}\sigma_s$, the Green function $G_{2,\mathrm{im}}$ being scale invariant and the background potentials transforming as $V_{\text{bg},1}\rightarrow\lambda V_{\text{bg},1}$ and $V_{\text{bg},3}\rightarrow\lambda^3V_{\text{bg},3}$. The derivative in the second line can equally be computed as the work done against the electrostatic forces as the charges are displaced by the rescaling, using the total potential we have\footnote{We are assuming that the charge density vanishes at the tip, $\sigma_s\left(\rho_s\right)=0$, so that we do not encounter problems related to its divergence.}
\begin{align}
\begin{aligned}
    &\frac{d\mathcal{E}}{d\lambda}\Big|_{\lambda=1}\\
    &=\sum_s\Omega_2\int d\rho\,\rho^2\,\sigma_s\left(V'_{\hat s}(\rho,z_s) z_s+\underbrace{{\dot V}_{\hat s}(\rho,z_s)}_{=0}\rho\right)\\
    &=-\sum_s z_s F_s.
\end{aligned}
\end{align}
Therefore, we end up with
\begin{align}
    \sum_s z_s F_s=2U-W_{\text{ext},1}-3W_{\text{ext},3}
\end{align}
We now want to evaluate $W_{\text{ext},i}$ explicitly and to trade $U$ for a more convenient quantity. For the latter, consider a uniform rescaling of the charges\footnote{Again, such a rescaling does not in general give rise to equipotential disks away from $t=1$. However, as above, redistributing the charges within each disk does not affect the energy variation at linear order, and we only need the first variation at $t=1$.}, $\sigma_s\rightarrow t\sigma_s$. Denoting by $\Phi=V-V_{\text{bg}}$ the potential generated by the conductors and their images, the total potential becomes $V_{\text{bg}}+t\Phi$, since the image charges scale together with the physical ones. The work done in adding charge gives
\[
    \frac{d\mathcal{E}}{dt}=\sum_s\int_{D_s}d^3x\,\sigma_s\left(V_{\text{bg}}+t\Phi\right),
\]
where $D_s$ denotes the $s$-th disk. Integrating from zero charge to $t\sigma_s$, one obtains
\begin{align}
    \begin{aligned}
    \mathcal{E}[\{t&\sigma_s\};\{z_s\}]\\
    &=\frac{t^2}{2}\sum_s\int_{D_s}d^3x\,\sigma_s\Phi
      +t\sum_s\int_{D_s}d^3x\,\sigma_s V_{\text{bg}}\\
    &=t^2U+tW_{\text{ext}}.
    \end{aligned}
\end{align}
At $t=1$ the total potential is constant on each conductor, and the same charge variation gives
\begin{align}
   \frac{d\mathcal{E}}{dt}\Big|_{t=1}=2U+W_{\text{ext}}=\sum_s Q_s V_s.
\end{align}
Solving for $U$,
\begin{align}
    U=\frac{1}{2}\sum_s Q_s V_s-\frac{1}{2}W_{\text{ext}}
\end{align}
so that
\begin{align}
    \sum_s z_s F_s=\sum_s Q_s V_s-2W_{\text{ext},1}-4 W_{\text{ext},3}.
\end{align}
It remains to evaluate the work done against the background potential,
\begin{align}
&\begin{aligned}
    W_{\text{ext},1}=-&\sum_s\Omega_2\int d\rho\, \rho^2\,\sigma_s(\rho)\\
    &\frac{\eta\mu^3}{2}\left(\rho^2+z_s^2\right)^{1/2}U_1\left(\frac{z_s}{\sqrt{\rho^2+z_s^2}}\right),
\end{aligned}\\
&\begin{aligned}
    W_{\text{ext},3}=-\sum_s&\Omega_2\int d\rho\, \rho^2\sigma_s(\rho)\\
    &\frac{\mu^5}{2^9}\left(\rho^2+z_s^2\right)^{3/2}U_3\left(\frac{z_s}{\sqrt{\rho^2+z_s^2}}\right)
\end{aligned}
\end{align}
which are proportional to the dipole and octupole moments of the charge distribution, $P$ and $q_3$. In particular, with the normalization of the $U_\ell$ functions above we obtain $W_{\text{ext},1}=-\eta\mu^3P/2$ and $W_{\text{ext},3}=-\mu^5\,q_3/2^8$, so that 
\begin{align}
\begin{aligned}
    \sum_s z_s F_s&=\sum_s Q_sV_s+\eta\mu^3P+\frac{\mu^5}{2^6}q_3\\
    &=\sum_s Q_sV_s\Big|_{\eta=0}+\frac{\eta\mu^3}{2}P+\frac{\mu^5}{2^6}q_3.
\end{aligned}
\end{align}
This identity was used in \eqref{eq:total_bulk}.

\paragraph{Legendre transform from electrostatics} As explained in \autoref{sec:canonical}, the bundle data fix the potentials $V_s$ under continuous fluctuations in the path integral. Therefore, the grand-canonical on-shell action $I$ should be understood as a function of the conductors' potentials $\{V_s\}$ with the charges $\{Q_s\}$ given by the on-shell relation, i.e. solving the electrostatic problem at fixed potentials. In this paragraph we show that we can perform the Legendre transform trading $V_s$ for its conjugate variable explicitly using identities from electrostatics. The Legendre transform reads
\begin{align} \label{eq:Legendre}
    I_{\text{can}}=I-\sum_sV_s\frac{\partial I}{\partial V_s}
\end{align}
with $I_{\text{can}}$ understood as a function of $\{Q_s\}$, with the potentials $\{V_s\}$ given again by solving the electrostatics problem at fixed conductor charges. Using the results of the previous paragraph one can show that the on-shell action is proportional to the binding energy of the conductors
\begin{align}
    I=-\mathcal{K}U,
\end{align}
where $\mathcal{K}$ is defined in \eqref{eq:K}.
We now derive a standard relation from classical electrodynamics. In particular the electrostatic energy $\mathcal{E}$ can be written as
\begin{align}
\begin{aligned}
    \mathcal{E}&=U+W_{\text{ext}},\\
    &=\frac{1}{2}\int_{z>0} d^4x\,\left(\nabla\Phi\right)^2+\sum_s\int_{D_s}d^3x\,\sigma_sV_{\text{bg}}.
\end{aligned}
\end{align}
Here $\Phi$ is again the potential normalized by the background value, $\Phi=V-V_{\text{bg}}$. Let us now vary $U$ with respect to the disk potentials, keeping the background fixed. Integrating by parts and using Laplace equation away from the disks (where $V$ is differentiable)
\begin{align}
\begin{aligned}
    &\delta U
    =\int_{z>0} d^4x\,\nabla\Phi\cdot\nabla\delta\Phi\\
    &=\sum_s\Omega_2\int_0^{\rho_s} d\rho\,\rho^2
    \left(V'\right|_{z_s^-}-\left.V'\right|_{z_s^+})\delta V_s\\
    &=\sum_s Q_s\,\delta V_s,
\end{aligned}
\end{align}
where $\delta\Phi=\delta V_s$ on the disk because the background potential is fixed. There is a subtlety in this derivation: $\rho_s$ depends on the geometry and therefore on $\{V_s\}$, so in principle $\delta U$ could receive a contribution from the variation of $\rho_s$. However, the corresponding endpoint contribution is proportional to $\sigma_s(\rho_s)$, which vanishes by regularity. We therefore obtain
\begin{align}
    \frac{\partial U}{\partial V_s}=Q_s,
\end{align}
which matches \cite[eq. 2.7]{Landau:ECM}. We can plug this back in the Legendre transform equation \eqref{eq:Legendre},
\begin{align}
\begin{aligned}
    \sum_sV_s\frac{\partial I}{\partial V_s}=-\mathcal{K}\sum_sQ_sV_s
\end{aligned}
\end{align}
obtaining the same result as \eqref{eq:final} of \autoref{sec:canonical}, that was instead obtained through dualization of $C_2$. As a comment, we notice that
\begin{align}
    I_{\text{can}}=\mathcal{K}\mathcal{E},
\end{align}
proportional to the total electrostatic energy of the conductors in the background. This is expected since $V_s=\partial\mathcal{E}/\partial Q_s$, and it shows that the Legendre transform is involutive.